\documentclass[final, 5p, sort&compress, times, authoryear, twocolumn]{elsarticle}

\usepackage{graphicx}
\usepackage{wrapfig, framed, caption}
\usepackage[cmex10]{amsmath}
\usepackage{stackengine}

\usepackage{float}
\usepackage{amssymb}
\usepackage{siunitx}
\usepackage{hyperref}
\usepackage{cleveref}
\allowdisplaybreaks
\usepackage[toc,nopostdot,nonumberlist]{glossaries}
  
\usepackage{acro}
\makeglossaries
\usepackage[latin1]{inputenc}
\usepackage{tikz}
\usetikzlibrary{shapes,arrows}
\usepackage{multirow}
\usepackage[normalem]{ulem}
\useunder{\uline}{\ul}{}
\usepackage{algorithm,algcompatible,amsmath}
\usepackage{lineno}
\usepackage{multirow}

\modulolinenumbers[5]
\setcitestyle{square,sort,comma,numbers}
\journal{Energy}

\begin{document}
\begin{frontmatter}

\title{Multi-Stage Compound Real Options Valuation in Residential PV-Battery Investment}

\author[label1]{Yiju Ma\corref{cor1}}
\address[label1]{School of Electrical and Information Engineering, The University of Sydney, Sydney, Australia}

\cortext[cor1]{Corresponding author}

\ead{yiju.ma@sydney.edu.au}

\author[label1]{Kevin Swandi}
\ead{kevinswandi@gmail.com} 
 
\author[label1]{Archie C. Chapman}
\ead{archie.chapman@sydney.edu.au}

\author[label1]{Gregor Verbi\v{c}}
\ead{gregor.verbic@sydney.edu.au}

\begin{abstract}

Strategic valuation of efficient and well-timed network investments under uncertain electricity market environment has become increasingly challenging, because there generally exist multiple interacting options in these investments, and failing to systematically consider these options can lead to decisions that undervalue the investment. In our work, a \textit{real options valuation} (ROV) framework is proposed to determine the optimal strategy for executing multiple interacting options within a distribution network investment, to mitigate the risk of financial losses in the presence of future uncertainties. 
To demonstrate the characteristics of the proposed framework, we determine the optimal strategy to economically justify the investment in residential PV-battery systems for additional grid supply during peak demand periods. 
The options to defer, and then expand, are considered as multi-stage compound options, since the option to expand is a subsequent option of the former. 
These options are valued via the \textit{least squares Monte Carlo} method, incorporating uncertainty over growing power demand, varying diesel fuel price, and the declining cost of PV-battery technology as random variables. 
Finally, a sensitivity analysis is performed to demonstrate how the proposed framework responds to uncertain events. 
The proposed framework shows that executing the interacting options at the optimal timing increases the investment value.

\end{abstract}


\end{frontmatter}

\section{Introduction}
Peak demand is the time when consumer demand for electricity is at its highest. In distribution systems, the installed power transfer capacity must be greater than the expected annual peak demand. However, in reality, it has become common for forecast electricity demand to exceed the distribution network's supply capacity in the near, as peak demand continues to increase \cite{AEMO2}. 
In response to this, traditionally, distribution network service providers (DNSPs) have sometimes invested in new generation in substations (e.g. diesel generators) to accommodate the growing peak demand within their operation areas. Specifically, diesel generators make sense for very `peaky' peaks, i.e. seasonal/weather driven, characterized by high peak-to-average load profiles. However, this is an expensive method considering that the diesel costs are uncertain due to the relatively high capital cost and price uncertainty of liquid fuels. 

Residential PV-battery systems have become an attractive alternative for effectively providing peak demand support, reducing the risk of supply falling short of demand. One of the key drivers to this development is the falling cost of PV-battery systems with the fast advancement of the technology \cite{AEMO3}. Meanwhile, the benefits of residential PV-battery systems are well recognized. First, battery systems store surplus PV generation, and utilize the stored energy in the evening for peak-load support \cite{tonkoski2012impact}. Second, they reduce network power losses and help with voltage regulation \cite{ma2019novel}. Although the technology is currently expensive to implement, it has already been shown effective in reducing diesel generation \cite{scott2019network}, and thus providing opportunities to replace the costly diesel generator investment, as well as to defer expensive network augmentation \cite{Clean,Evan}. 

However, with the increasing uncertainties in distribution network investment and operation in the electricity system, DNSPs confront a great challenge in determining efficient and well-timed investment in residential PV-battery systems. In Australia and some other jurisdictions, asset ring-fencing regulations mean that these PV-battery systems cannot be owned by (regulated) DNSPs if they are also used for supplying energy and services to contestable markets. However, to promote the use of this technology, DNSPs may provide incentives to cover a large proportion of PV-battery procurement and installation cost, and in this way, they can effectively invest in these assets.

Given this context, the decision rule of the traditional \textit{net present value} (NPV) analysis is to undertake the investment immediately if the NPV is positive and reject those with a negative NPV, regardless of how future uncertainties will unfold \cite{vlaovic2013advantages}. However, this method ignores the \textit{real options} in an irreversible distribution network investment, such as the options to \textit{defer}, \textit{expand}, \textit{contract} and \textit{abandon}. Each of these options is a contingent decision that the investor has the flexibility to make based on the realization of future uncertainties. This highlights the importance and value of managerial flexibility, which is not captured in standard NPV assessments.

In contrast, in this paper we use \textit{real options valuations} (ROV) to identify the options embedded in an investment, and hence the possible flexible investment directions in the light of various uncertainties. Specifically, ROV takes into account the value of managerial flexibility of making contingent decisions upon the realizations of future uncertainties, typically captured by \textit{Monte Carlo} (MC) simulations. Given this, we reduce the exposure to risk, by considering management's right, but not obligation, to make contingent decisions \cite{kodukula2006project}. 

However, there are generally several interacting options embedded in an investment. The most common interaction happens when a subsequent option becomes available after the prerequisite option is executed. In this case, the subsequent option provides additional future contingency that affects the investor's decisions. 
Examples include system expansion, relocation, re-contracting and abandonment which often appear only after the option to invest has been executed. Properly incorporating these subsequent options into the valuation of the prerequisite option may lead to a different investment strategy, and therefore provide additional value to the investment. Nonetheless, there is currently limited literature that presents a valuation framework to explicitly consider the interdependency among the options available in a distribution network investment. 

\subsection{Contributions}
Against this background, this work proposes a multi-stage valuation framework that uses ROV to evaluate an investment with compound options under multiple future uncertainties. To demonstrate the characteristics of the framework, we apply it to assess the economic benefits and costs accrued to a DNSP for providing incentives to network customers to purchase PV-battery systems. 
By doing so, we remove the need to install additional substation diesel generators. 

Given this, we consider two interacting options in this investment, which are:
\begin{itemize}
  \item The option to \textit{defer} the PV-battery investment in the first decision period; and
  \item The option to \textit{expand} this investment in the second decision period.
\end{itemize}

The option to expand is a subsequent option, which is only considered after the option to defer is executed. Given this, the value of the deferral option is also dependent on this subsequent option. The proposed framework derives the optimal investment strategy subject to these options by quantitatively taking into account the uncertainties and the flexibility of making decisions contingent to unfolding information. 

The uncertainties considered in this work are:
\begin{itemize}
  \item Growing peak power demand which decides the capacity of the PV-battery investment, as well as diesel generator operational cost;
  \item Varying diesel price that determines the cost of diesel procurement; and
  \item The declining cost of PV-battery technology.
\end{itemize}

These uncertainties are chosen as the state variables to the ROV because their future projections affect the behaviour of the cash flow, and the stochastic variations provide opportunities for a DNSP to increase the investment value by making decisions contingent on their realizations.

In order to capture the value of the options, and determine the optimal investment strategy within a pre-identified decision period, the framework overcomes two main challenges that have not been addressed in the existing literature: (i) the need to enable ROV on the application of PV-battery hybrid system investment and determine the optimal investment timing given multiple interacting options available, under multiple uncertainties,  and (ii) integrate multiple investments (diesel generator and PV-battery systems) within one ROV process to raise the potential for greater option values. Thus, the major contributions of the proposed framework are summarized as follows:
\begin{itemize}
  \item We define the economic benefits of the PV-battery investment as the cost saving from replacing the expensive diesel generation investment to enable the application of ROV;
  \item We use ROV to determine the optimal strategy that subjects to multiple interacting options available in the application of PV-battery system investment; and
  \item By combining the investments in diesel generator and PV-battery systems, we integrate the uncertainties involved in both investments within one valuation process to provide greater opportunity values.
\end{itemize}

By doing so, we demonstrate the process of deriving the optimal strategy for a distribution network investment that has multiple interacting options. Although ROV based on the LSMC method has already been used in the evaluation of network investments, this is the first time that the usefulness of these techniques has been shown for assessing the investment value of customer-owned PV-battery systems in particular, and for distribution investments in general.

Our results show that, from the traditional NPV analysis, the PV-battery investment is abandoned immediately because the NPV is negative. However, this decision is changed after considering the options embedded in the investment. Specifically, by considering the option to expand in the future, the proposed ROV framework suggests to delay the investment and thereby increase the value of the investment. In addition to this, we observe that changing the drift parameter and volatility of the state variables can have significant impacts on the distribution of optimal strategies.   
\subsection{Literature review}
In this subsection we review existing literature for: (i) ROV in transmission and distribution network investments, and (ii)~the methods to calculate the value of real options.

In recent years, ROV has been frequently applied for valuation of distribution and transmission network investments, including transmission network expansion \cite{salazar2007decision, pringles2015real}, distributed generation \cite{zou2017optimal}, and renewable generation including hydro-power \cite{linnerud2014investment}, wind generation \cite{boomsma2012renewable, bockman2008investment, eryilmaz2016does,ritzenhofen2016optimal,boomsma2015market}, solar generation \cite{martinez2013assessment, zhang2016real, gahrooei2016timing, zhang2017optimal, eissa2017lobatto} and large scale battery storage systems \cite{ma2019estimating}. A common feature of these studies is that they determine the value to execute one or several \textit{independent} options embedded in a network investment in the presence of uncertain electricity market conditions or regulatory policies, and hence the optimal investment timing. 

However, in practice, a distribution network investment often involves multiple \textit{interacting} (i.e.~not independent) options which actively engage with each other to produce a greater investment value. For this setting, the authors in \cite{loncar2017compound} determine the value of multiple options, including the option to invest, followed by the subsequent options (expand, re-power, contract and abandon), in a wind farm investment. However, the investment timing for the subsequent options are fixed (i.e.~a European call option) to reduce the size of the problem. 
Currently, the method to properly value American compound options, with flexible execution timing, has not been presented in the existing literature. 
More importantly, we have found that the topic of addressing investment in hybrid renewable generation systems, such as PV-battery, using ROV has not been presented, which drives our research direction. 

Given this context, we need to decide the most suitable approach to address the problems above. 
The \textit{partial differential equations} (PDE) approach \cite{de2006flexibility, macdougall2015value} based on the research of Black, Scholes and Merton, and \textit{binomial lattice} \cite{eryilmaz2016does, boomsma2015market, eissa2017lobatto} have been widely applied to calculate the value of real options. However, the PDE approach can be used to incorporate only one uncertainty, or at most two correlated ones. This is not the case in the electricity market where multiple uncertainties exist. Meanwhile, PDE is designed for European-type of option analysis, in which the investment can only be taken on a specific future date \cite{schachter2016critical}. On the other hand, lattice models use backward induction, and the prior path of the underlying variable is unknown at the time computations are made, making it impossible to incorporate multiple interacting options with numerous state variables \cite{schachter2016critical}. Thus, these methods are not suitable for distribution network investments where there are multiple sources of uncertainty and interacting options. 

In contrast, \textit{binomial tree} models can handle the cases with compound options under multiple uncertainties better, by adding additional decision nodes \cite{loncar2017compound}. However, their computation complexity grows exponentially with the increasing number of decision nodes, constraining their use to complex but smaller-sized problems. 

In light of this, the \textit{least square Monte Carlo} (LSMC) method \cite{longstaff2001valuing} is applied to determine the value of real options, and hence derive the optimal investment strategy \cite{pringles2015real, zou2017optimal, linnerud2014investment, boomsma2012renewable, zhang2016real, zhang2017optimal, blanco2011real}. This method allows us to incorporate many sources of uncertainty, and accurately capture the flexibility in delaying investments using MC simulations. For example, the authors in \cite{blanco2011real} applied this method to value flexible AC transmission systems devices in transmission networks, by modeling both demand and fuel costs as stochastic processes, and providing the optimal timing for the option to install, locate and remove the asset. 
For these reasons, the LSMC method is used in our work. 

The rest of the paper is organized as follows. Section \ref{section2} explains the LSMC approach which is used for valuing the managerial flexibility. Section \ref{section3} describes the costs and benefits analysis of the PV-battery investment, and presents the stochastic modeling of future uncertainties including power demand growth, varying diesel fuel price, and the declining cost of PV-battery systems. The outcomes are evaluated in Section \ref{section4}. Finally, Section \ref{section5} draws conclusions. 

\section{Real Options Theory} \label{section2}
The traditional NPV method fails to appraise an investment under uncertainties and in the presence of contingency, as it considers managerial flexibility as a passive factor and attains only deterministic investment decisions. In contrast, ROV recognizes the benefits of contingency and includes this as an active entity in its calculation, which potentially changes the value of the investment \cite{kodukula2006project}.

For multi-stage compound options valuation, we need to determine the value of the subsequent options (the option to expand), and incorporate this value when valuing the deferral option. In this case, the option to defer is to be executed within a 5-year decision period, $t \in \mathcal{T}_\mathrm{inv}$, while the option to expand is considered in the next 5-year decision period, $t \in \mathcal{T}_\mathrm{exp}$. The optimal investment strategy is extracted from this process. The mathematical formulation for solving compound options via the LSMC method is detailed in this section. 

\subsection{ROV with the LSMC method}
To determine the optimal investment strategy subject to the compound options, the first task is to apply the LSMC method to calculate the value of the subsequent option to expand the investment, assuming that the investment has already been carried out. Specifically, this method combines a forward-looking model for incorporating uncertainties, and a backward recursion (least square regression) for determining the value of an option \cite{schachter2016critical}. The option can be executed any year from the initial year $t_0$ to the maturity $T_{\mathrm{m}}$. This time-span is divided in to $n \in \mathcal{N}$ intervals, whose length is 1 year.

We assume that there are $h\in \! \mathcal{H}$ options within an investment, thus, we use $h$ and $h\! +\! 1$ to represent the options to defer and expand, respectively. Given this, the investment value in year ${t}$ considering the option to expand is denoted as ${F_{h+1}(t_{n},{X}_{t_{n}})}$, where ${{X}_{t_{n}}}$ is the state variable of the investment, including the growing power demand, varying diesel fuel price, and the declining PV-battery cost. Hence, the future discounted value of the subsequent investment option can be expressed as follows:
\begin{equation} \label{ROVeq1}
F_{h+1}(t_{0},{X}_{t_{0}}) = \max_{\tau \in [t_{0}, {T}_{\mathrm{m}}]}\left\{e^{-r(\tau-t_{0})}E^{*}_{t_{0}}\left[\Pi_{h+1}(\tau,X_{\tau})\right]\right\},
\end{equation} 
\noindent where $\tau$ is the optimal stopping time on each MC path, $\Pi_{h+1}(\tau,X_{\tau})$ is the payoff for expanding the project, ${{E}^{*}_{t_{n}}\left[\Pi_{h+1}(\tau,X_{\tau})\right]}$ is the expectation on the information available at ${t_{n}}$, and ${r}$ is the risk neutral discount rate.

LSMC integrates MC simulations with least square regression to accurately estimate the option value. The dynamic features of state variables are simulated by generating a set, ${\Omega}$, of MC realization paths by means of \textit{geometric Brownian motion}. Then, we estimate the continuation value, denoted ${\Phi(t_{n},\omega,X_{t_{n}})}$, which is essentially an estimate of the investment value of the next time step, given a realization $\omega \in \Omega$. This value represents the value of continuing to wait for the realization of future random variables at each time along the ${\omega^{\mathrm{th}}}$ path. By comparing ${\Phi(t_{n},\omega,X_{t_{n}})}$ to the value of investing immediately, the optimal stopping time along each MC path is found. The process for calculating the optimal stopping times is described by the \textit{Bellman's principle of optimality}, which is expressed as follows:
\begin{equation}\label{ROVeq2}
F_{h+1}(t_{n},{X}_{t_{n}}) = \max\left\{\Pi_{h+1}(t_{n},X_{t_{n}}),\Phi_{h+1}(t_{n},X_{t_{n}})\right\},
\end{equation}
\noindent where
\begin{equation}\label{ROVeq3}
\Phi_{h+1}(t_{n},X_{t_{n}}) = E^{*}_{t_{n}}\left[\sum_{i=n+1}^{N}e^{-r(t_{i}-t_{n})} \Pi_{h+1}(t_{n},t_{i},\tau)\right].
\end{equation} 
 
In more detail, the least square regression estimates the continuation value by regressing the discounted future investment values on a linear combination of a group of basis functions of the current state variables. Each group of the basis functions represents the payoff trajectory within a time interval, and we use these basis functions to estimate the payoff at $t\! + \! 1$ (continuation value). This process has been employed in many recent studies, which generally apply simple powers of the state variable ${{X}_t}$ as basis functions \cite{zou2017optimal}, \cite{blanco2011real}. Given this, we define the orthonormal basis of the ${j^\mathrm{th}}$ state variable as ${L_{j}}$. The optimal coefficients, ${\hat{\phi}_{j}}$, for the basis functions are obtained using \eqref{ROVeq4}. 
\begin{equation}\label{ROVeq4}
	\begin{split}
		\begin{aligned}
			\hat{\phi}_{j}(t) & = \underset{\phi_{j}(t)}{\operatorname{argmin}}  \biggl\{\sum_{i=n+1}^{N} e^{-r(t_{i}-t_{n})} \Pi_{h+1}\left(t_{i},X_{t_{i}}(\omega)\right) \\ & \quad - \sum_{j=1}^{J}\phi_{j}(t)L_{j}\left(X_{t_{n}}(\omega)\right)\biggl\}^{2}.
		\end{aligned}
	\end{split}
\end{equation} 
The continuation value for each MC path is thus calculated by feeding the optimal coefficient ${\hat{\phi}_{j}}$ into the linear combination of the basis functions, that is:
\begin{equation}\label{ROVeq5}
\Phi_{h+1}\left(t,X_{t_{n}}(\omega)\right) = \sum^{J}_{j=1} \hat{\phi}_{j}(t_{n})L_{j}\left(X_{t_{n}}(\omega)\right).
\end{equation}
The option value is maximized along each path if the option is executed as soon as the payoff exceeds the continuation value. The optimal stopping time along each of the in-the-money paths is determined by applying Bellman's principle of optimality, given by \eqref{ROVeq2} recursively from maturity ${{T}_{\mathrm{exp}}}$ to ${{t}_{0}}$. If the decision rule holds true at ${t_{n}}$, the stopping time ${\tau_{\omega}}$ along the ${{\omega}^{\mathrm{th}}}$ path will be updated to ${t_{n}}$, that is:
\begin{equation}\label{ROVeq6}
\text{if} \ \ \Phi_{h+1}\left(t_{n},X_{t_{n}}(\omega)\right) \leq \Pi_{h+1}\left(\tau,X_{t_{n}}(\omega)\right), \ \ \text{then} \ \ \tau(\omega) = t_{n}.
\end{equation} 
The optimal stopping time of each MC path ${\tau_{\omega}}$ forms a unique optimal stopping time matrix, which includes the earliest investment timing for all in-the-money MC paths. Using this matrix, the option value at $t_{0}$ that considers managerial flexibility and future uncertainty is determined by the following equation:
\begin{equation}\label{ROVeq7}
	\begin{split}
		\begin{aligned}
	       F_{h+1}(t_{0},{X}_{t_{0}}) &= \frac{1}{|\Omega|} \sum\limits_{\omega\in\Omega}{ e^{-r\tau(\omega)}\Pi_{h+1}\left(\tau(\omega),X_{\tau(\omega)}(\omega)\right) }.
		\end{aligned}
	\end{split}
\end{equation} 
Given the value of the subsequent option, to calculate the value of the option to defer the PV-battery investment, the Bellman's principle of optimality described by \eqref{ROVeq2} becomes: 
\begin{equation}\label{ROVeq8}
	\begin{split}
		\begin{aligned}
			F_{h}(t_{n},{X}_{t_{n}}) &= \max\left\{\Pi_{h}(t_{n},X_{t_{n}})+F_{h+1}(t_{n},X_{t_{n}}), \Phi_{h}(t_{n},X_{t_{n}})\right\},
		\end{aligned}
	\end{split}
\end{equation} 
\noindent where 
\begin{equation}\label{ROVeq81}
\Phi_{h}(t_{n},X_{t_{n}}) = E^{*}_{t_{n}}\left[\sum_{i=n+1}^{N}e^{-r(t_{i}-t_{n})} \sum_{l=h}^{H} \Pi_{l}(t_{n},t_{i},\tau)\right].
\end{equation} 
For the deferral option value, the continuation value, $\Phi_{h}(t,X_{t})$ is compared to the payoff of the investment, $\Pi_{h}(t,X_{t})$, plus the value of the option to expand, $F_{h+1}(t,X_{t})$. Thus, the stopping time ${\tau_{\omega}}$ along the ${{\omega}^{\mathrm{th}}}$ path is updated to ${t}$, if:
\begin{equation}\label{ROVeq9}
	\begin{split}
		\begin{aligned}
             \Phi_{h}\left(t_{n},X_{t_{n}}(\omega)\right) \leq \left(\Pi_{h}\left(t_{n},X_{t_{n}}(\omega)\right) +F_{h+1}\left(t_{n},X_{t_{n}}(\omega)\right)\right).
		\end{aligned}
	\end{split}
\end{equation} 
We use the updated decision rule in the LSMC method to calculate the deferral option value, and hence the optimal investment strategy. The accuracy of the estimation grows as the number of simulation paths and basis functions increases. The presented method can be applied to multiple subsequent options, where the payoff  of the $(h+n)^{\mathrm{th}}$ option needs to be incorporated when calculating the value of the $(h+n-1)^{\mathrm{th}}$ option in \eqref{ROVeq8}, \eqref{ROVeq81} and \eqref{ROVeq9} within the LSMC approach.

\section{Costs and Benefits Analysis} \label{section3}

To carry out the financial assessment of the PV-battery investment, we need to determine the cost and payoff for carrying out and expanding the investment, respectively. These values are used as the inputs to the ROV to value the compound options, and therefore the corresponding optimal investment strategy. The valuation algorithm is described in Algorithm \ref{alg1}, and discussed in more detail below.

Specifically, we calculate the cost of the investment via the traditional NPV method, and hence the payoff ($\Pi_{h,t,\omega}$) from replacing the diesel generator investment with the PV-battery investment during the first 5-year decision period, and the payoff ($\Pi_{h+1,t,\omega}$) from expanding this investment in the next 5-year decision period. This period is chosen so that the DNSP can fully capture the benefit of delaying the PV-battery investment, while carrying out the appropriate investment for peak demand support within a relatively short time-span. The detailed formulation is described in this section. Then, we demonstrate the modelling of future uncertainties including (i) growing power demand, (ii) varying diesel fuel price, and (iii) declining cost of the PV-battery technology.

\subsection{Payoff Calculation}

The energy delivered through the transformer that is over the thermal limit for each day between 2014 and 2017 is calculated from the aggregated historical electricity usage data from the Top Ryde substation (\SI{132}{\kilo\volt}/\SI{33}{\kilo\volt}), which has a thermal limit of \SI{35}{MVA}. These data, provided by Ausgrid\footnote{a DNSP in New South Wales, Australia; see
\url{https://www.ausgrid.com.au/Industry/Innovation-and-research/Data-to-share/Distribution-zone-substation-data}.}, have a 15-minute resolution. The future growth in power demand that is over the thermal limit is then simulated. We take the average across the simulated data for each future decision year, and use it as the aggregated installed capacity for the PV systems and diesel generator, denoted as $E^{\mathrm{Cap}}_{t,\omega}$. The battery size is decided based on the PV size. In Australia, \SI{2}{kWh} of battery is typically used per \SI{1}{kW} of PV installed. 

The future costs of the PV-battery investment ($c^{\mathrm{PV}}_{t,\omega}$) and diesel generator investment ($c^{\mathrm{DG}}_{t,\omega}$) need to be discounted via the traditional NPV method as follows before calculating the payoffs: 
\begin{equation} \label{eq3}
	\begin{aligned}
		NPV = \sum_{t=1}^{T}{\frac{c_t}{(1+r)^t}}. 
	\end{aligned}
\end{equation}
\noindent where $r$ is the risk-free discount rate\footnote{In our work, we assume that the future costs are risk free. Thus, we have fixed this value to a constant (0.06).}.

We assume that the DNSP is responsible for 70\% of PV-battery procurement (the rest is paid by the customers), and it is obliged to cover the cost of electricity usage over the thermal limit and is not covered by the additional generation capacity provided by PV-battery systems (denoted as ${c^{\mathrm{g}}_{t,\omega}}$). This cost occurs when the PV-battery investment is delayed or the peak demand exceeds the installed generation capacity. Thus, ${\Pi_{h,t,\omega}}$, including ${c^{\mathrm{g}}_{t,\omega}}$ is given by:    
\begin{equation} \label{eq1}
	\begin{aligned}
		\Pi_{h,t,\omega} = \left(c^{\mathrm{PV}}_{t,\omega} - c^{\mathrm{DG}}\right) E^{\mathrm{Cap}}_{t,\omega} - c^{\mathrm{OM}} + c^{\mathrm{g}}_{t,\omega}, 
	\end{aligned}
\end{equation} 

\noindent where $c^{\mathrm{PV}}_{t,\omega}$ and $c^{\mathrm{DG}}$ are the costs per \SI{1}{kW} for the same aggregated capacity $E^{\mathrm{Cap}}_{t,\omega}$, $c^{\mathrm{OM}}_{h,t,\omega}$ is the maintenance and operation cost, while $c^{\mathrm{g}}_{t,\omega}$ is given by:
\begin{equation} \label{eq2}
	\begin{aligned}
		c^{\mathrm{g}}_{t,\omega} = c^{\mathrm{f}}_{t,\omega}\Delta{E^{\mathrm{g}}_{t,\omega}},
	\end{aligned}
\end{equation}
\noindent where $c^{\mathrm{f}}_{t,\omega}$ is the diesel price, and $\Delta{E^{\mathrm{g}}_{t,\omega}}$ is the electricity usage over the thermal limit and fails to be covered by PV-battery systems.

\begin{table}[t]
\centering\small

\caption{Cost specifications.}
\begin{tabular}{cccc}
\hline
Items                                               & Cost\\
PV-battery system, $c^{\mathrm{PV}}_{t,\omega}$     & Risk neutral valuation     \\
Diesel generator, $c^{\mathrm{DG}}$                 & {\$}{600}/kW \cite{el2005integrated}             \\
Peak demand, $E^{\mathrm{Cap}}_{t,\omega}$          & GBM                       \\
Operation and maintenance, $c^{\mathrm{OM}}$        & {\$}{100}{k/year} \cite{el2005integrated}         \\
Diesel fuel, $c^{\mathrm{f}}_{t,\omega}$            & Mean-reverting process    \\
\hline
\end{tabular}
\label{T1}
\end{table}

The aggregated capacity of the PV-battery systems installed in the first decision period does not cover the power demand growth in the future, which leaves an open question as to whether the expansion is necessary. If the option to expand is abandoned, the future power demand growth will be covered by installing an additional diesel generator. 

In more detail, the payoff for expanding the PV-battery investment, ${\Pi_{h+1,t,\omega}}$, assuming that the investment has already been executed, is the difference between the cost of installing additional PV-battery systems to cover the peak demand growth after the first decision period, and the cost of expanding the diesel generation investment for the same amount of aggregated generation capacity, $E^{\mathrm{Cap}}_{h+1,t,\omega}$; that is:
\begin{equation} \label{eq_payoff2}
	\begin{aligned}
		\Pi_{h+1,t,\omega} = \left(c^{\mathrm{PV}}_{t,\omega} - c^{\mathrm{DG}}\right) E^{\mathrm{Cap}}_{h+1,t,\omega} - c^{\mathrm{OM}}_{h+1,t,\omega} + c^{\mathrm{g}}_{t,\omega}. 
	\end{aligned}
\end{equation} 

In our work, we assume that the DNSP is in charge of electricity distribution, network planning, monitoring and maintenance. The electricity bills generated from the diesel generator are allocated to this entity, and these are the primary source of cash inflow in this investment. Cash outflows in this investment include system procurement, installation, maintenance and operation costs for the generator.


\begin{algorithm}[t]
\footnotesize
    \caption{Real options valuation algorithm}
  \begin{algorithmic}[1]
  	\STATE Generate $\Omega$ realization paths of $\Delta{E^\mathrm{Cap}}$, $c^{\mathrm{f}}$ and $c^{\mathrm{PV}}$.
    \FOR{$\omega$ = 1:$\Omega$}
        \FOR{year = 6:10}
        	\STATE Calculate $\Pi_{h+1}$. 
        \ENDFOR     
    \ENDFOR
    
    \FOR{$t$ = $T_{\mathrm{exp}}$:-1:1}
    	\STATE Calculate $\Phi_{h+1}$ via LSMC for all in-the-money paths.  
        \FOR{$\omega$ = 1:$\Omega$}
        	\IF{${\Pi} >= \Phi_{h+1}(t,X_{t}(\omega))$}
            	\STATE Update $\tau$ = t.            	
            \ENDIF
        \ENDFOR
    \ENDFOR
    \STATE Calculate the $F_{h+1}$ using \eqref{ROVeq7}.
    \STATE Repeat the steps 3 to 6 to calculate $\Pi_{h}$ from Years 1 to 5

    \FOR{$t$ = $T_{\mathrm{inv}}$:-1:1}
    	\STATE Calculate $\Phi_{h}$ via LSMC for all in-the-money paths.  
        \FOR{$\omega$ = 1:$\Omega$}
            \STATE Update the decision rule to \eqref{ROVeq9}.
        	\IF{$\Pi + F_{h+1}>= \Phi_{\mathrm{h}}(t,X_{t}(\omega))$}
            	\STATE Update $\tau$ = t.            	
            \ENDIF
        \ENDFOR
    \ENDFOR
    \STATE Calculate $F_{h}$ and extract the optimal investment strategy.
    
  \end{algorithmic}
  \label{alg1}
\end{algorithm}

\subsection{Modelling of Future Uncertainties} \label{SectionGBM}
In this subsection we simulate the random variables used in the payoff calculation, including growing power demand that exceeds the thermal limit ($\Delta{E^{\mathrm{Cap}}_{t,\omega}}$), cost of diesel fuel ($c^{\mathrm{f}}_{t,\omega}$), and the declining cost of PV-battery technology ($c^{\mathrm{PV}}_{t,\omega}$). Specifically, the growing peak demand that exceeds the thermal limit ($\Delta{E^{\mathrm{Cap}}_{t,\omega}}$) is governed by the GBM and simulated using MC analysis. GBM is used in this case as we assume that the stochastic peak demand evolution over time can be captured by a GBM process. This assumption relies on the fact that the increments of process in the GBM show the Markov property, which assumes that any future change is independent from the previous values, while the variable remains positive throughout the process \cite{fleten2007optimal}. The GBM for the growing peak demand is therefore described mathematically as follows:
\begin{equation} \label{eq4}
	\begin{aligned}
		dS^{\mathrm{d}}_{t} = \mu S^{\mathrm{d}}_{t}dt + \sigma S^{\mathrm{d}}_{t} d W_{t},
	\end{aligned}
\end{equation} 

\noindent where ${S^{\mathrm{d}}_{t}}$ describes the sought stochastic process of power demand, ${\mu}$ is the percentage drift that describes the rate of growth in the aggregated peak demand that is over the thermal limit, ${\sigma}$ is the percentage volatility of the data, and ${W_{t}}$ is the Wiener process that describes the stochastic component. Thus, the discretization recursion formula is given by:
\begin{equation} \label{eq5}
	\begin{aligned}
		S^{\mathrm{d}}_{t+\Delta{t}} = S^{\mathrm{d}}_{t}e^{\left(\mu-\frac{\sigma^2}{2}\right)\Delta{t}+\sigma dW_{t}}. 
	\end{aligned}
\end{equation} 

\begin{figure}[t]  
\centering
    \begin{framed}
    \includegraphics[width=7cm,keepaspectratio]{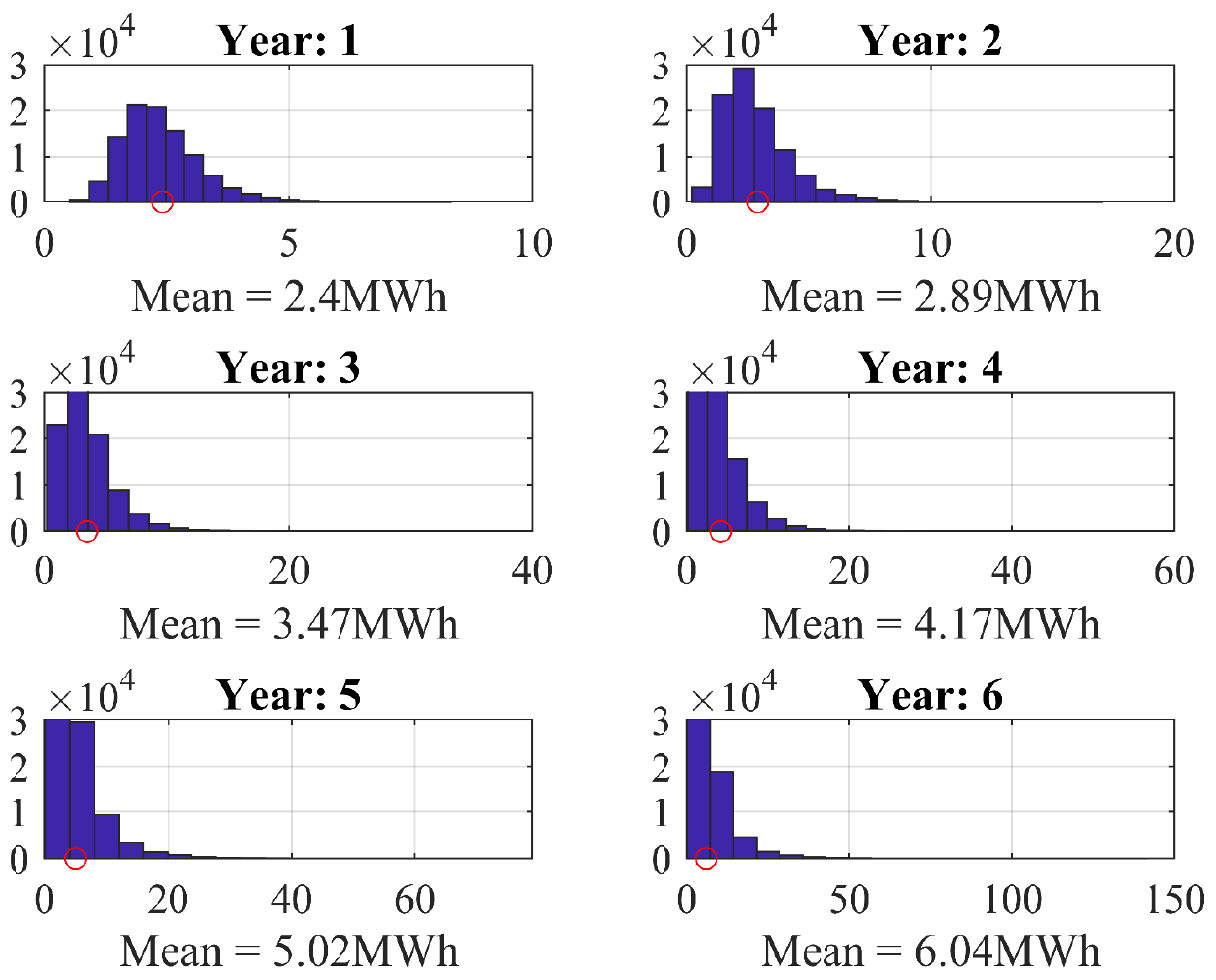}%
    \end{framed}
    \vspace{-8pt}
    \caption{Distribution of simulated average monthly growth in future peak demand that is over the transformer thermal limit via GBM (drift = 1.5\% and volatility = 9.8\%)}
    \label{fig11}
\end{figure}

\begin{figure}[t]
\centering
    \begin{framed}
    \includegraphics[width=7cm,keepaspectratio]{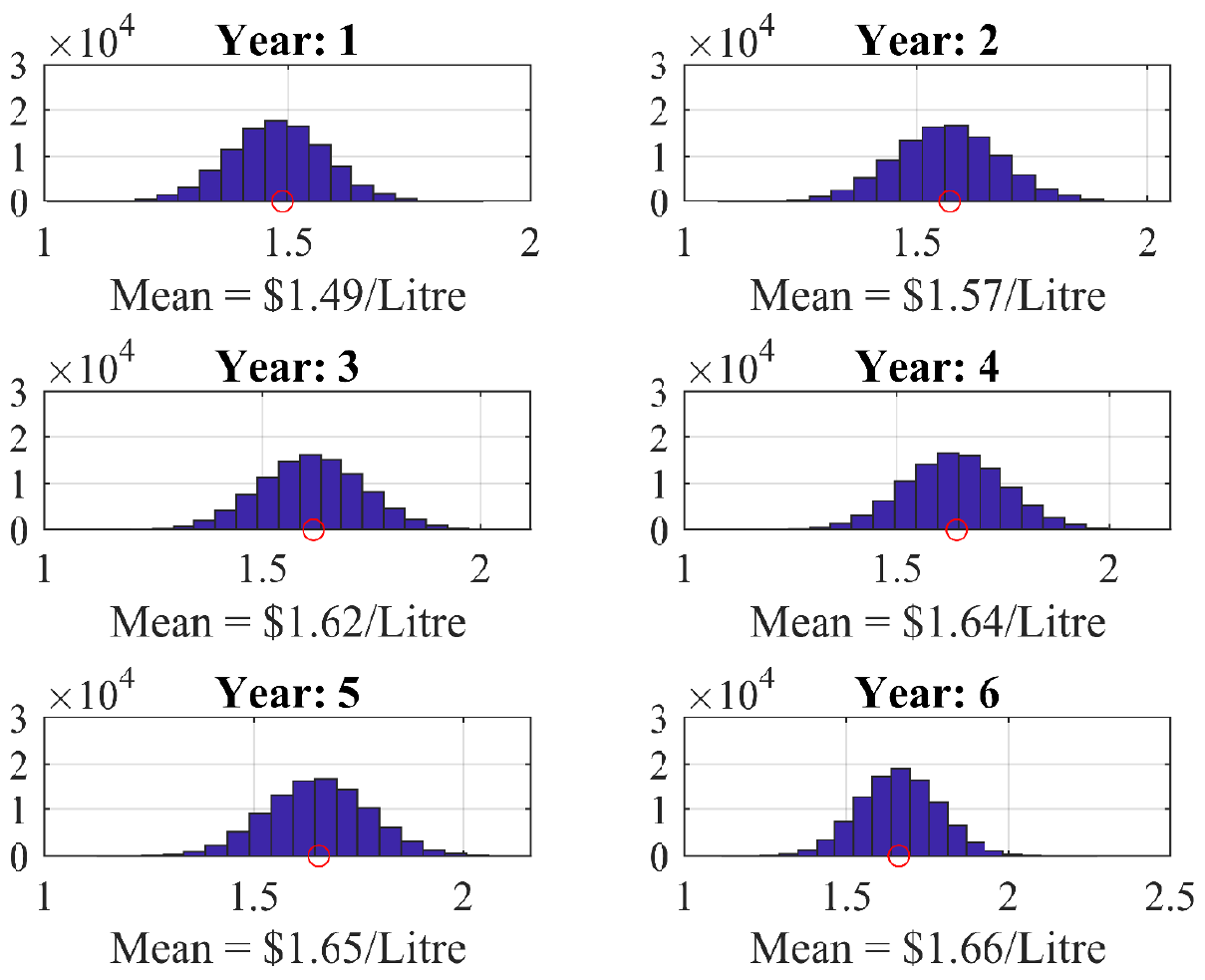}%
    \end{framed}
    \vspace{-8pt}
    \caption{Distribution of simulated future diesel price via a mean reverting process (speed of reversion = 5\%, reversion level = 2.6, volatility = 4.7\%)}
    \label{fig22}
\end{figure} 

\begin{figure}[t]
\centering
    \begin{framed}
    \includegraphics[width=7cm,keepaspectratio]{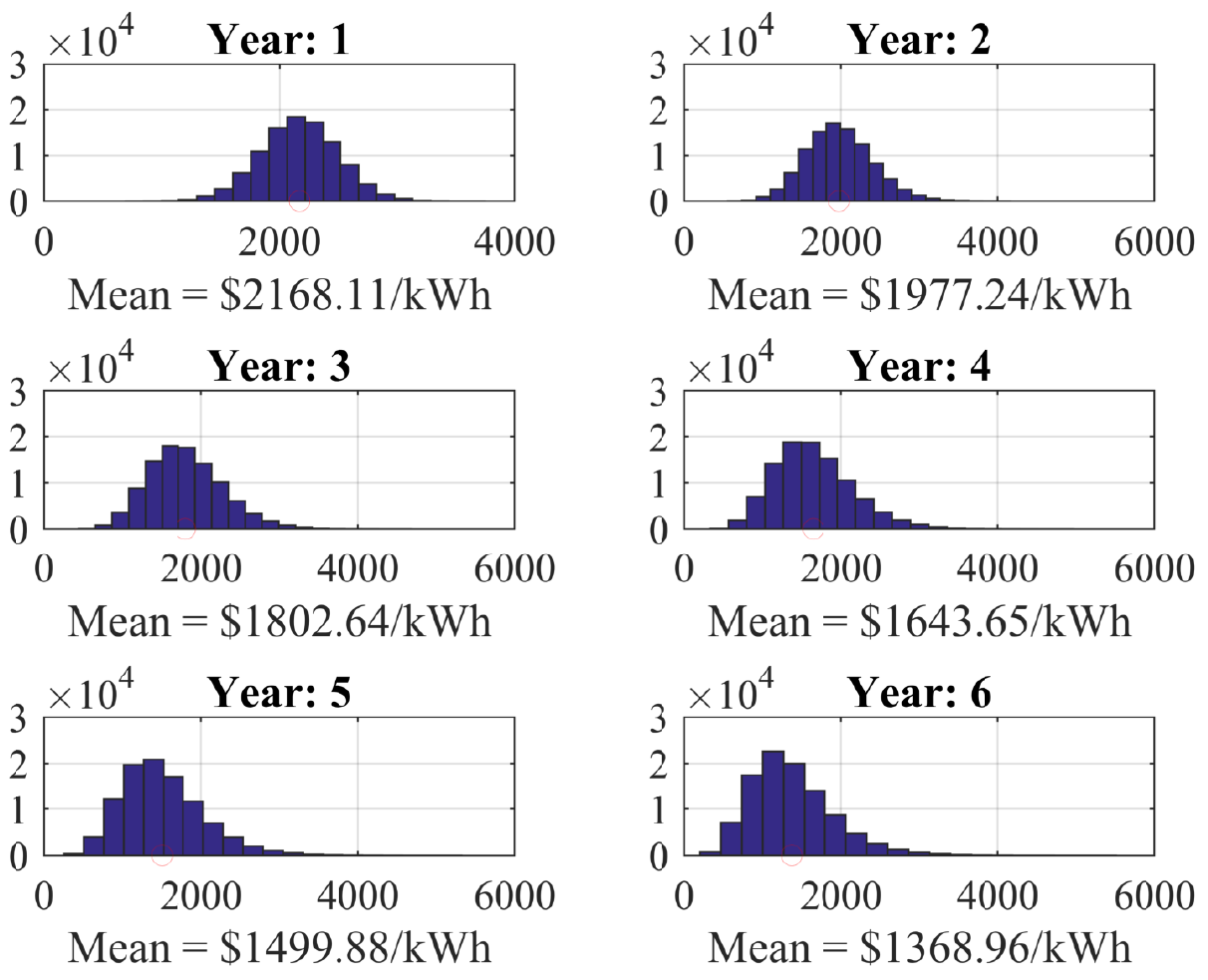}%
    \end{framed}
    \vspace{-8pt}
    \caption{Distribution of simulated future cost of PV-battery system via risk neutral valuation (risk-free rate = 0.06, volatility = 9\%)}
    \label{fig33}
\end{figure}

The simulation of future power demand takes the historical electricity usage data provided by Ausgrid. The future power demand is simulated for 10,000 MC paths. 

On the other hand, diesel fuel price, like all the other bulk commodities such as oil, gas and metal tends to conform to its long-term mean, with stochastic shocks when unforeseen "events" occur. Given this we used a mean-reverting process to simulate the varying diesel fuel prices; that is:

\begin{equation} \label{eqMRP1}
	\begin{aligned}
		dS^{\mathrm{f}}_{t} = \beta_{\mathrm{f}}(\hat{S}^{\mathrm{f}} - S^{\mathrm{f}}_t)dt + \sigma S^{\mathrm{f}}_{t} d W_{t},
	\end{aligned}
\end{equation} 

\noindent where ${S^{\mathrm{f}}_{t}}$ describes the sought stochastic process for the cost of diesel fuel, ${\beta_{\mathrm{f}}}$ is the speed of reversion to the mean, and $\hat{S}^{\mathrm{f}}$ is the reversion level. Thus, the discretization recursion formula is given by:

\begin{equation} \label{eqMRP2}
	\begin{aligned}
		S^{\mathrm{f}}_{t+\Delta{t}} = e^{-\beta_{\mathrm{f}}\Delta t}(S^{\mathrm{f}}_t - \hat{S}^{\mathrm{f}}) + \hat{S}^{\mathrm{f}} + \sigma \epsilon \sqrt{(1- e^{-2\beta_{\mathrm{f}} \Delta t })/2\beta_{\mathrm{f}}}.
	\end{aligned}
\end{equation} 

In the LSMC approach, we assume that the sample paths of costs of assets (PV-battery systems) over the relevant time horizon are simulated according to the risk-neutral measure\footnote{This is because these costs depend crucially on their risk as investors typically demand more profit for bearing more risk. It is difficult to adjust the simulated expected values based on an investor's preference, and therefore, risk neutral measure is used.}. Given this, we first describe the evolution of the cost using GBM:

\begin{equation} \label{eqRNM}
	\begin{aligned}
		dS^{\mathrm{PV}}_{t} = \mu S^{\mathrm{PV}}_{t}dt + \sigma S^{\mathrm{PV}}_{t} d W_{t},
	\end{aligned}
\end{equation} 

\noindent where $S^{\mathrm{PV}}_{t}$ describes the sought stochastic process for the cost of PV-battery systems. Then, we define $dW_t = d\hat{W}_t - \frac{\mu - r}{\sigma} dt$, where $d\hat{W}_t$ is the Brownian motion under risk neutral measure and $r$ is the risk-free rate. Substituting this to \eqref{eqRNM} yields the risk neutral valuation: 

\begin{equation} \label{eqRNM2}
	\begin{aligned}
		dS^{\mathrm{PV}}_{t} = r S^{\mathrm{PV}}_{t}dt + \sigma S^{\mathrm{PV}}_{t} d \hat{W}_{t}.
	\end{aligned}
\end{equation} 

Thus, the discretization formula becomes:

\begin{equation} \label{eqRNM3}
	\begin{aligned}
		S^{\mathrm{PV}}_{t+\Delta{t}} = S^{\mathrm{PV}}_{t}e^{\left(r-\frac{\sigma^2}{2}\right)\Delta{t}+\sigma d\hat{W}_{t}}. 
	\end{aligned}
\end{equation}

The historical PV and battery price and diesel price data are obtained from \cite{battery} and \cite{bureau}, respectively. For brevity, in this work, only 6 years of the simulated future data are shown. It is observed that the electricity usage that is over the thermal limits, $\Delta{E^{\mathrm{Cap}}_{t,\omega}}$, and the cost of diesel fuel $c^{\mathrm{f}}_{t,\omega}$ increase gradually in the future, as illustrated in Fig.~\ref{fig11} and Fig.~\ref{fig22}, respectively, while the cost of PV-battery $c^{\mathrm{PV}}_{t,\omega}$ continues to decrease (Fig.~\ref{fig33}).

\section{Case Studies} \label{section4}
In this section, we demonstrate that the proposed ROV framework can be used to evaluate a distribution network investment with interacting options under multiple uncertainties. Specifically, we value the PV-battery investment considering two interacting options, which are (i) the option to defer in the first 5-year decision period, and (ii) the option to expand the investment in the next 5-year decision period. The subsequent option for the investment expansion can only be considered after the investment has been carried out. The option values are calculated via the LSMC approach described in Section \ref{section2}. As previously described, the expansion of the PV-battery investment aims to cover the further growth in the aggregated peak power demand after the first decision period.  

The study period is 10 years, including the two consecutive 5-year decision periods, and the annual risk-neutral discount rate is assumed to be \SI{6}{\%}. The DG investment and PV-battery investment are thereafter referred to as P1 and P2, respectively.

\subsection{Diesel Generator Cash Flow Analysis}
The NPV for executing P1 in Year 1 is calculated in this subsection. The capital cost of diesel generator is {\$}{600} per \SI{1}{kW} \cite{el2005integrated}, while the future electricity generation cost is dependent on the increase in the aggregated peak demand and diesel fuel price simulated using the GBM in Section \ref{section3}. 

The average NPV of P1 decreases from {--\$}{900}{k} in Year 1 to greater than {--\$}{2}{M} by the end of the 10-year study period, as illustrated by Fig.~\ref{fig2}, left, this is due to the expenses in maintenance and electricity generation. Specifically, the magnitude of the outliers outside the boxes, especially below, increases significantly with respect to time, showing the increasing risks for large investment costs in the case of fast-growing peak demand and diesel fuel cost. We use the NPV of P1 in Year 10 in \eqref{eq1} to calculate the payoff, $\Pi_{h,t,\omega}$, of P2.

\subsection{PV-Battery Cash Flow Analysis}
\begin{figure}[t]%
    \centering
    \includegraphics[width=4cm,keepaspectratio]{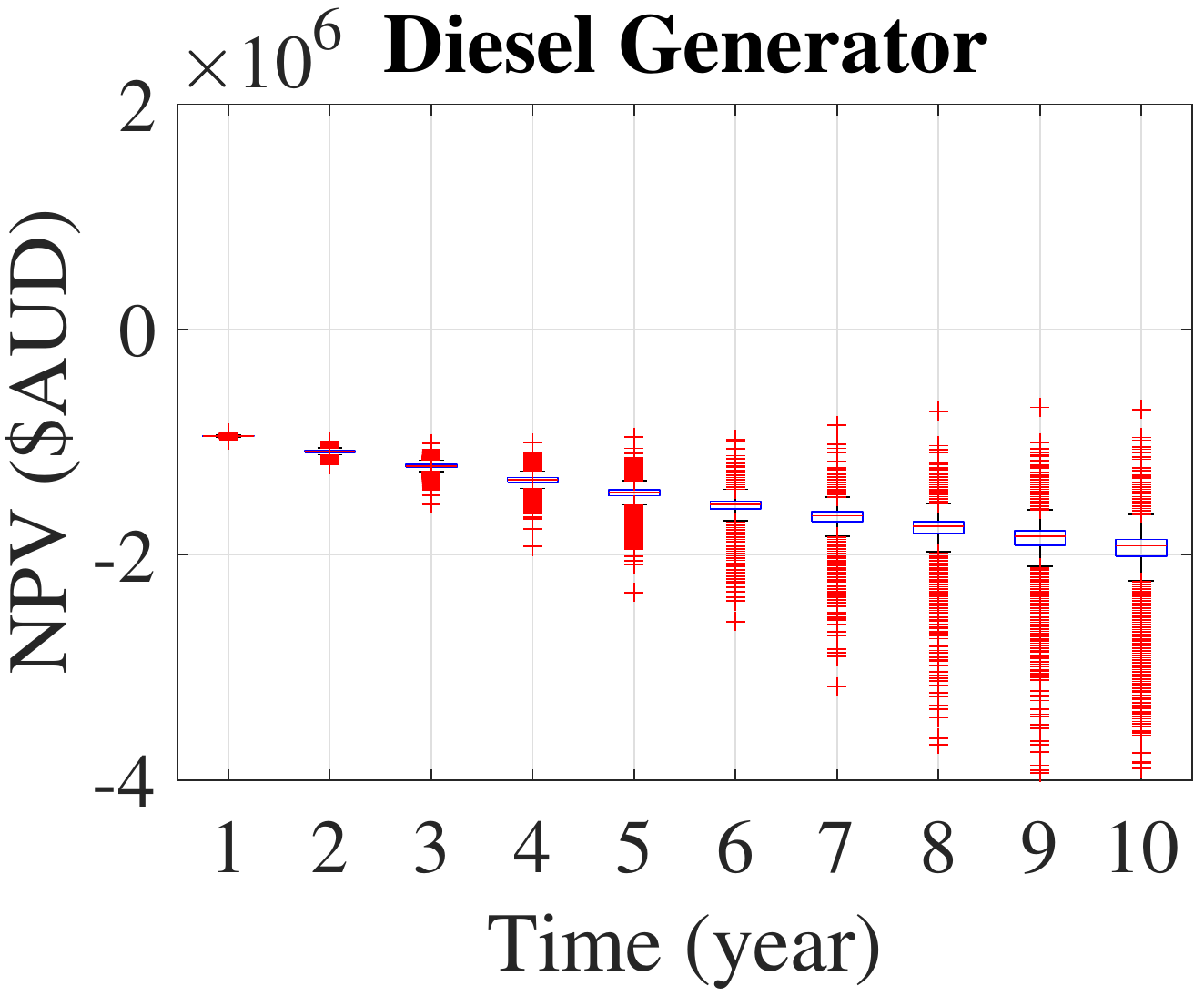}
    \qquad
    \includegraphics[width=4cm,keepaspectratio]{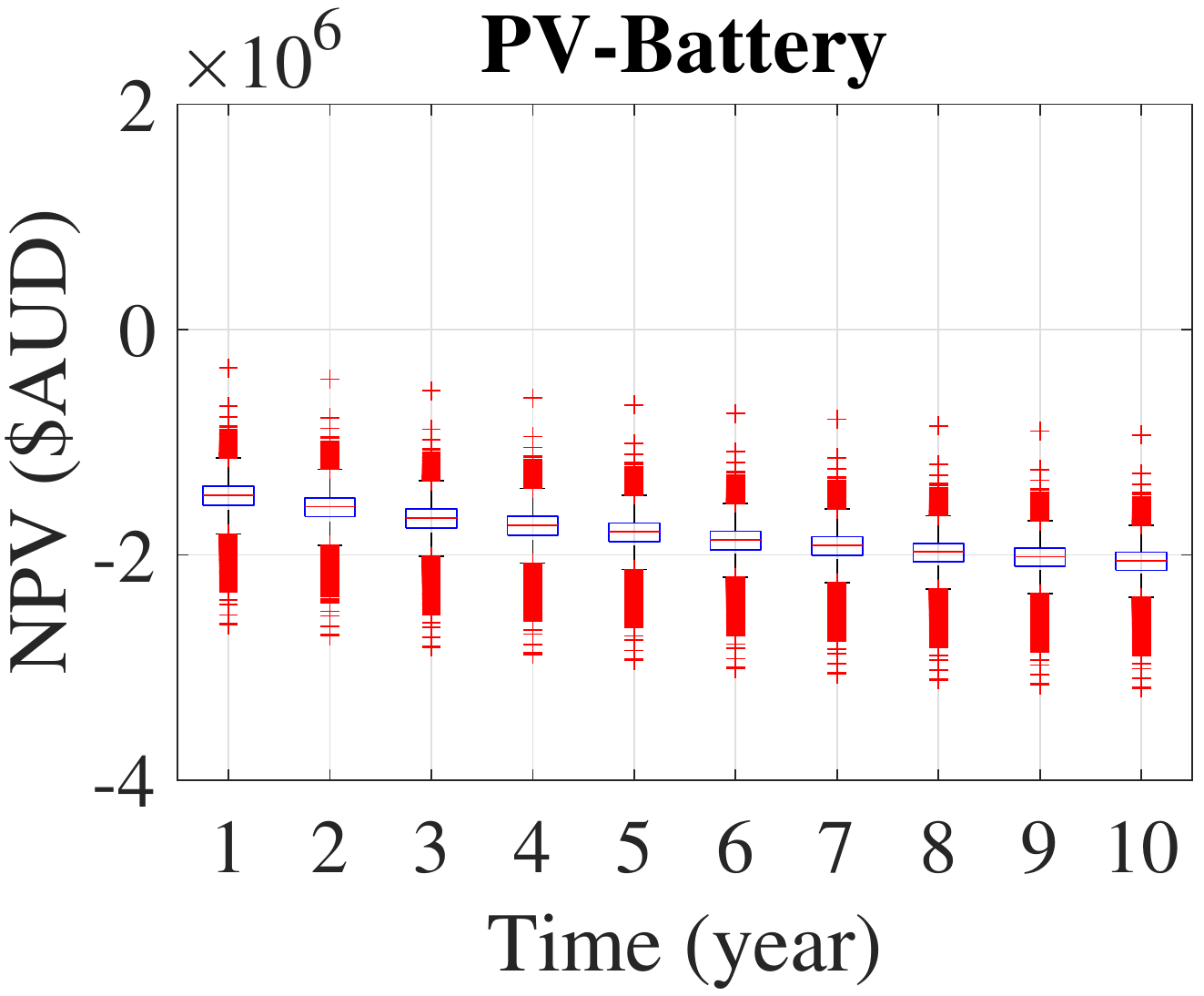}
    \caption{Costs for executing the diesel generator investment (left) and the PV-battery investment in Year 1 (right)}%
    \label{fig2}%
\end{figure}

Compared to the NPV of P1, the average NPV of P2 in Year 1 is roughly {--\$}{1.6}{M}, which decreases to just under {--\$}{2}{M} in Year 10, assuming P2 is executed in Year 1, as shown in Fig.~\ref{fig2}, right. The reduction in the NPV is less pronounced because PV-battery systems rely on cost-free solar power, and only the maintenance cost is committed to this investment during the study period. 

Based on the traditional NPV analysis, P2 is not profitable as the NPV in all MC paths are negative, as shown in Fig.~\ref{fig2}, right, and hence this investment is abandoned. However, using the ROV, we create an opportunity to make profits of this investment by postponing it to a time when the uncertainties turn favourable, and the trade-off to this is the cost to cover the electricity usage that is over the transformer thermal limit, given that neither P1 nor P2 is implemented until the optimal investment timing. 

\begin{figure}[t]%
    \centering
    \includegraphics[width=4cm,keepaspectratio]{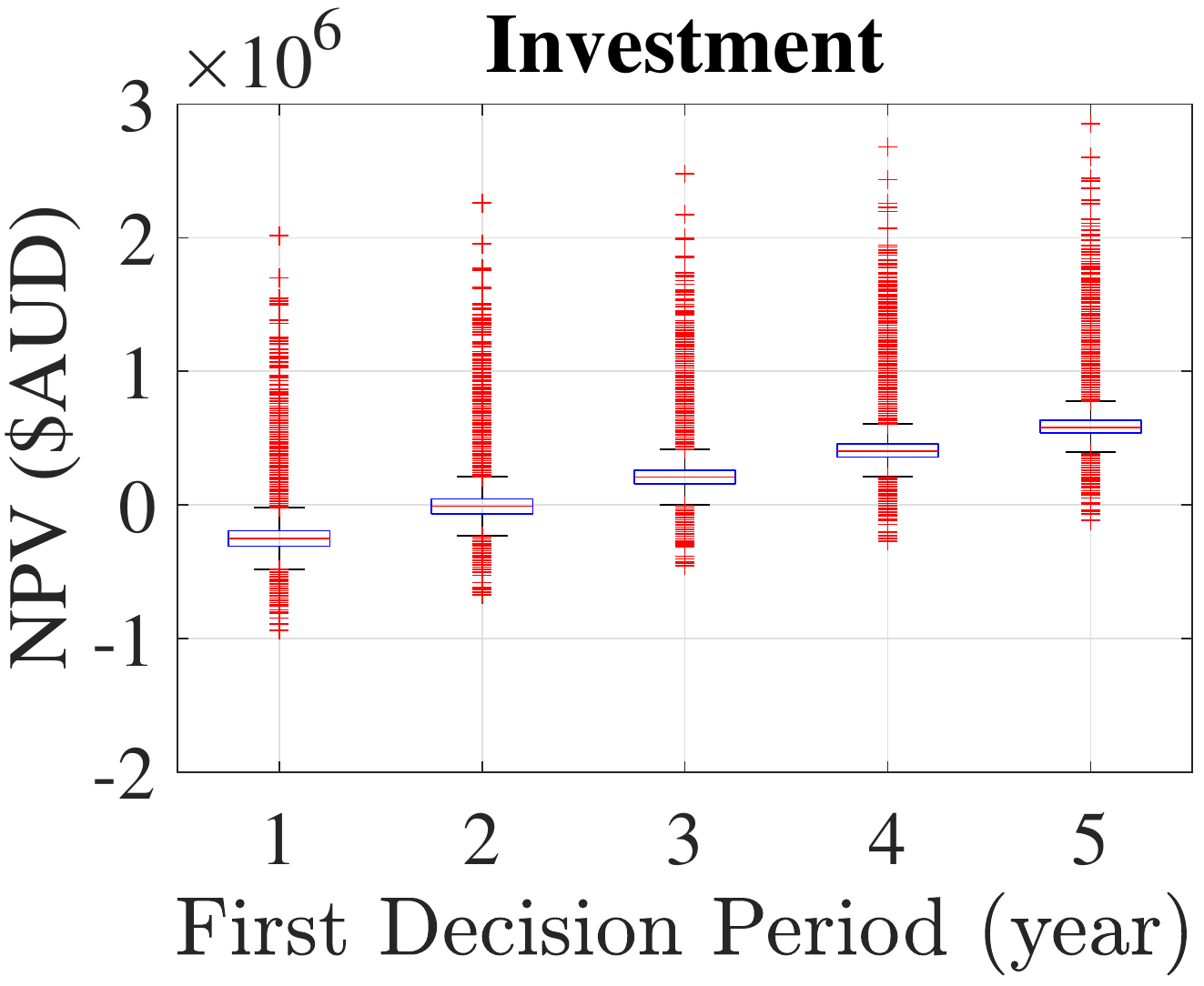}
    \qquad
    \includegraphics[width=4cm,keepaspectratio]{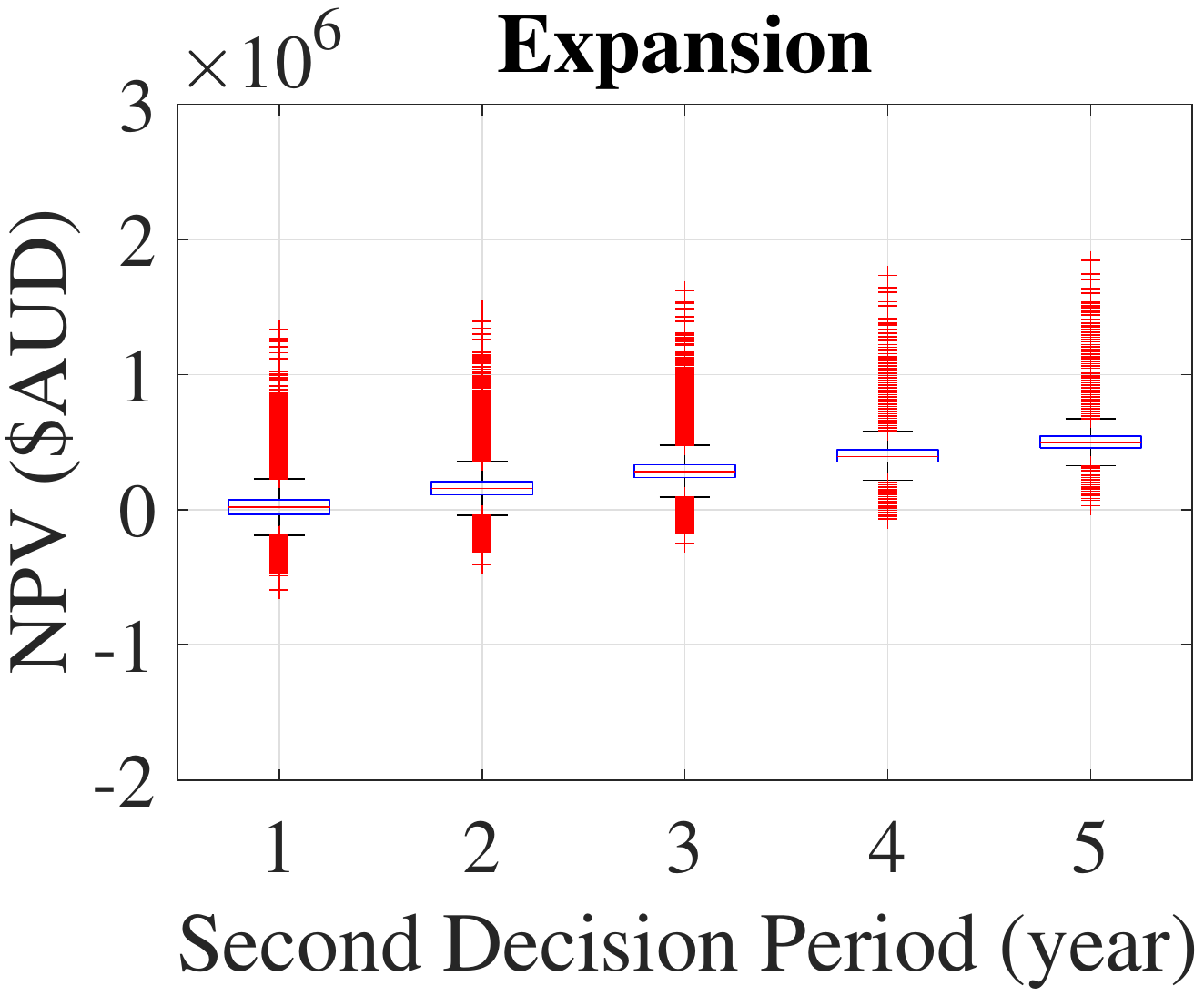}
    \caption{Payoffs for executing the PV-battery investment in each of the decision years (left), and for expanding the investment in each of the decision years (right)}%
    \label{fig3}%
\end{figure}

\begin{figure}[t]
	\centering
	\includegraphics[width=7cm,keepaspectratio]{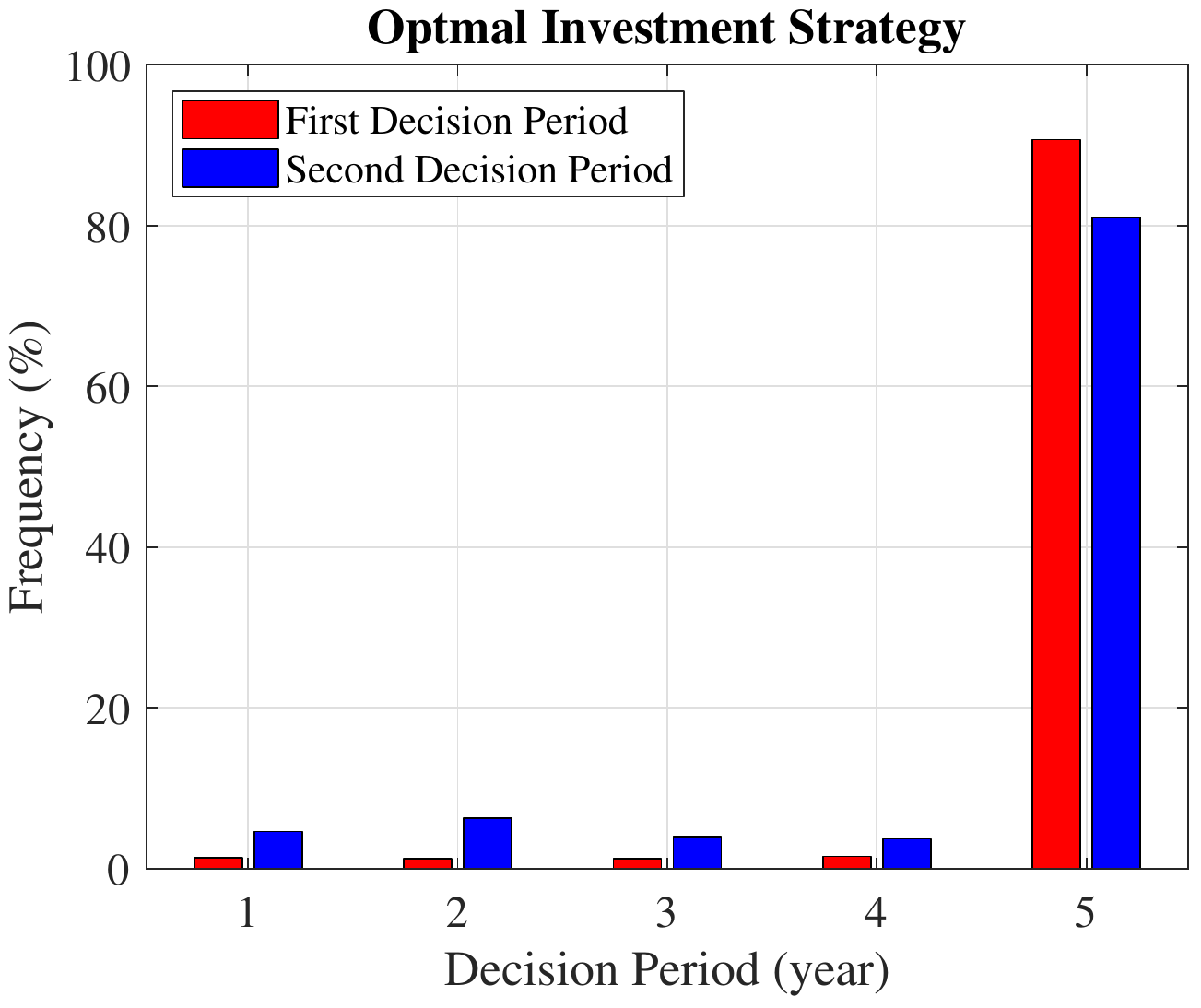}
	\caption{Frequency distribution of optimal investment timing for the option to defer the investment in the first decision period, and the option to expand the investment in the second decision period (benchmark).}
	\label{fig4}	
\end{figure}

The future payoffs of investing and expanding the PV-battery investment needed for the LSMC approach are shown in Fig.~\ref{fig3}. Specifically, these costs are calculated using \eqref{eq1} and \eqref{eq_payoff2}, respectively, where the cost parameters are shown in Table~.\ref{T1} and the simulation of the state variables is described in Section \ref{SectionGBM}. Observe that the payoffs for both investing in and expanding in the future gradually increase throughout the decision periods due to the declining cost of the advancing technology. Thus, there exists an opportunity to make the investment profitable in the future via ROV.

\begin{table*}[]
\footnotesize
\caption{ROV results under different scenarios.}
\begin{tabular}{ccccccccccc}
\hline
Scenario            & Description & Year 1 (\%) & Year 2 (\%) & Year 3 (\%) & Year 4 (\%) & Year 5 (\%) & Standard NPV (k\$) & ROV (k\$) & Flexible NPV (k\$) \\
S1                  & Benchmark  & 1.2  & 1.2   & 2.3  & 3.4  & 90            & --240                 & 660          & 420                                    \\
S2 & $\mu_{\mathrm{d}} = 3\%$  & 8.7 & 3.1 & 2.9 & 5.4 & 77 & --110                 & 740          & 630                   \\
S3 & $\sigma_{\mathrm{d}} = 20\%$ & 4.5  & 4.8 & 5.5 & 4.3 & 80 & --240                 & 600          & 360                   \\
S4 & $\beta_{\mathrm{f}} = 15\%$    & 2.2 & 0.5 & 0.3 & 0.4 & 97 & --230                 & 690          & 460                   \\
S5 & $\sigma_{\mathrm{f}} = 20\%$ & 0.3 & 6.1 & 4.8 & 5.0 & 84 & --250                 & 640          & 390                   \\
S6 & $\sigma_{\mathrm{pv}} = 20\%$  & 1.7 & 1.6 & 1.7 & 1.9 & 92 & --240                & 720          & 480                  \\
\hline
\end{tabular}
\label{T2}
\end{table*}


\begin{figure*}[t]%
    \centering
    \includegraphics[width=4cm,keepaspectratio]{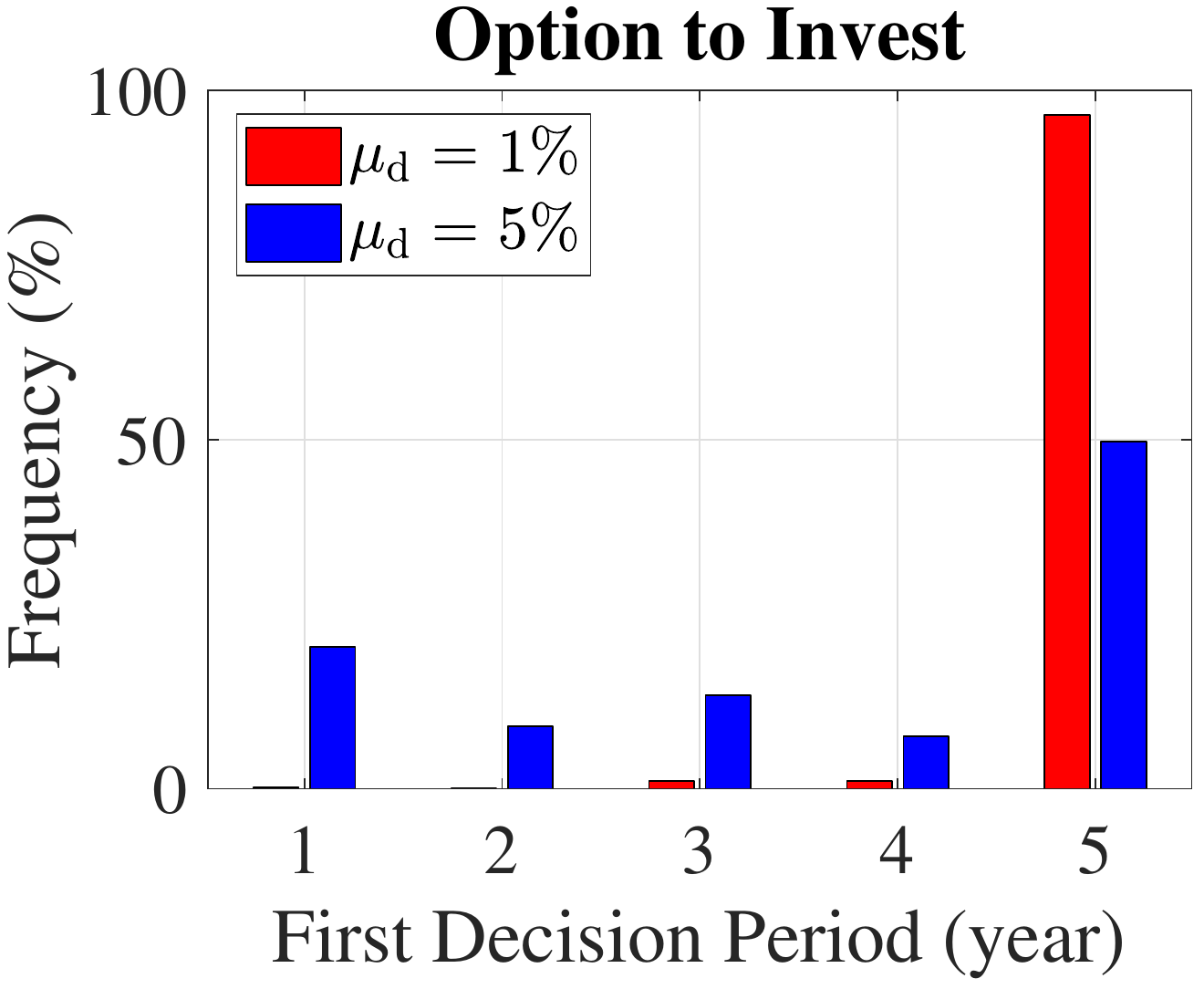}%
    \qquad
    \includegraphics[width=4cm,keepaspectratio]{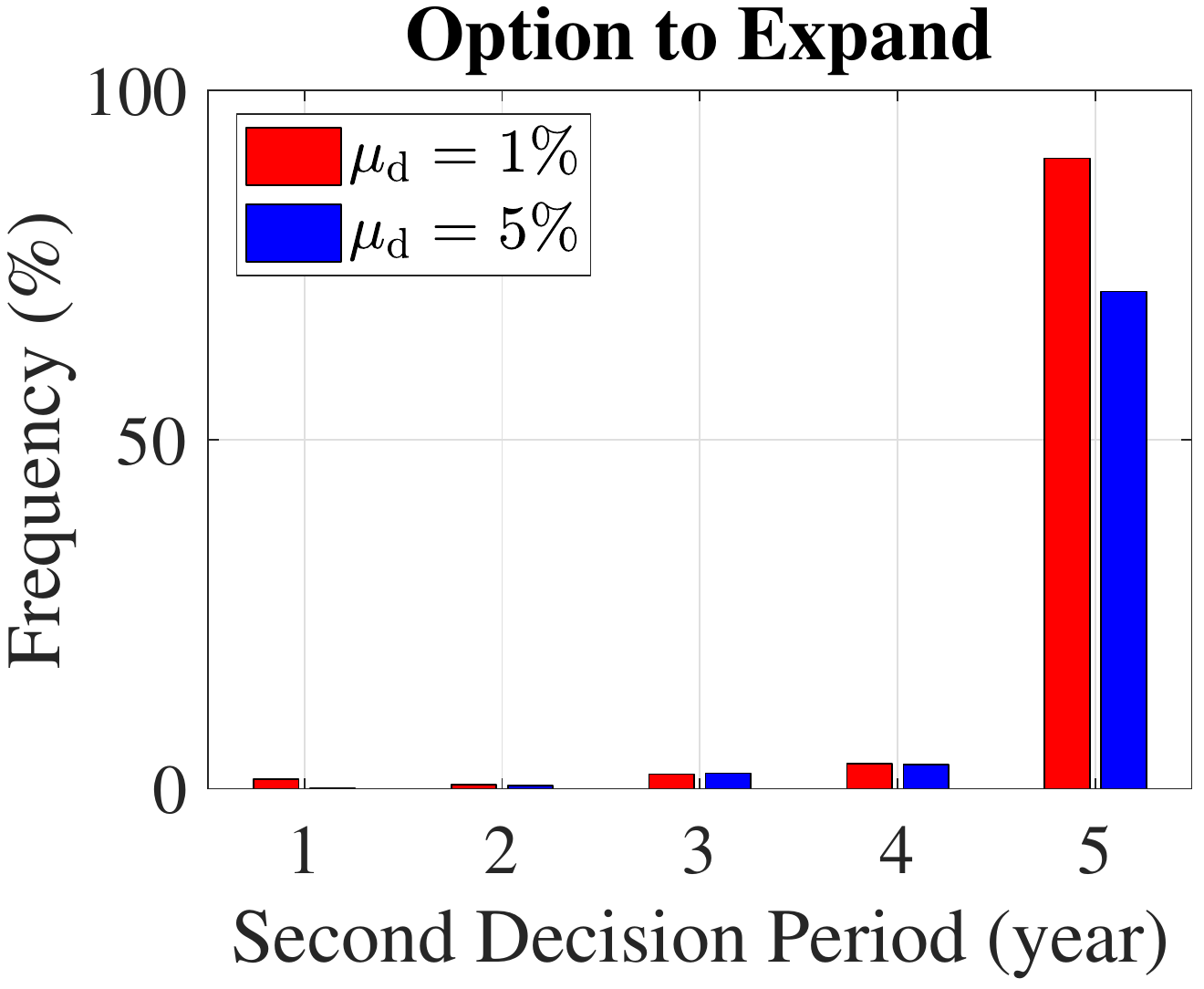}%
    \qquad
    \includegraphics[width=4cm,keepaspectratio]{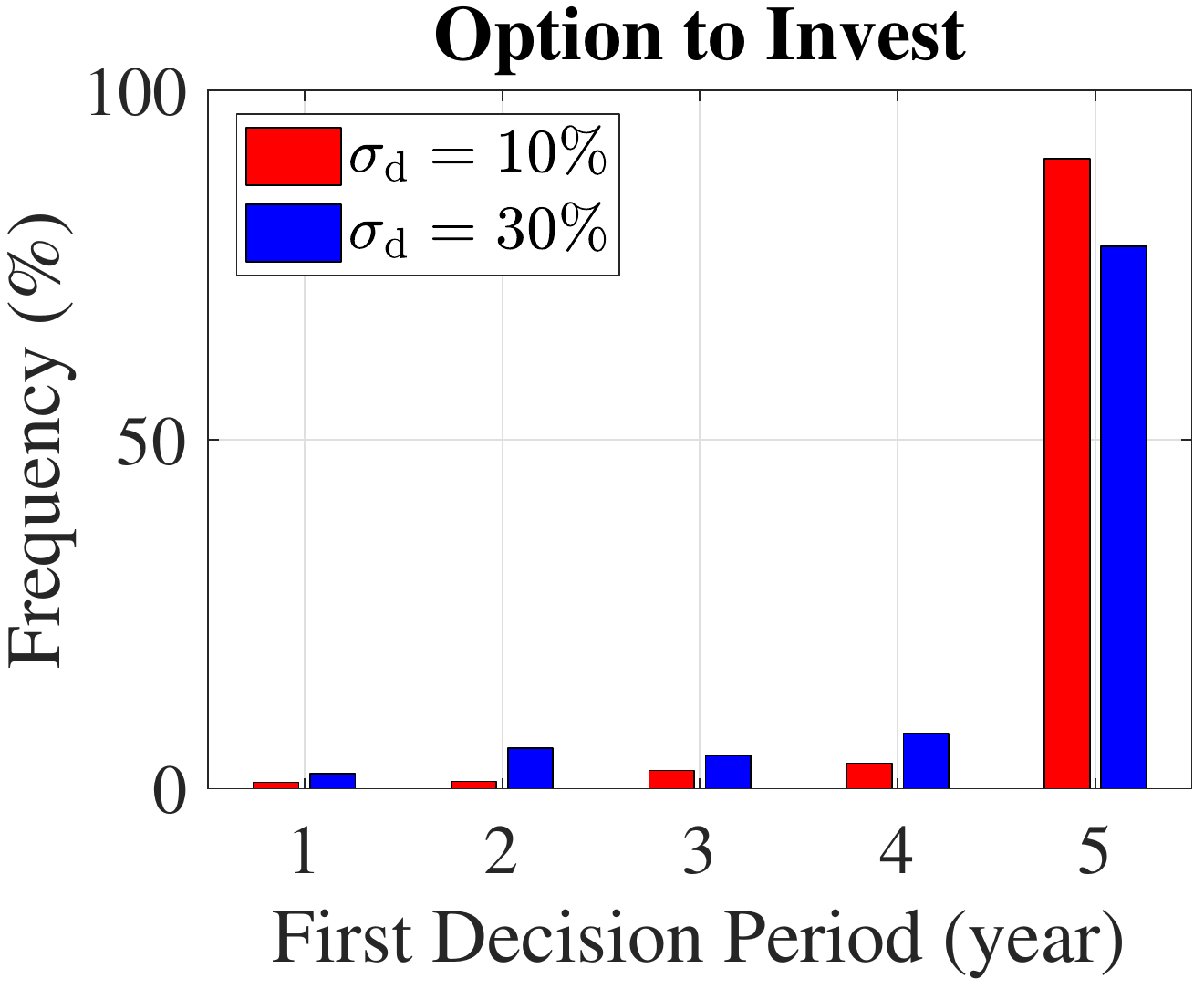}%
    \qquad
    \includegraphics[width=4cm,keepaspectratio]{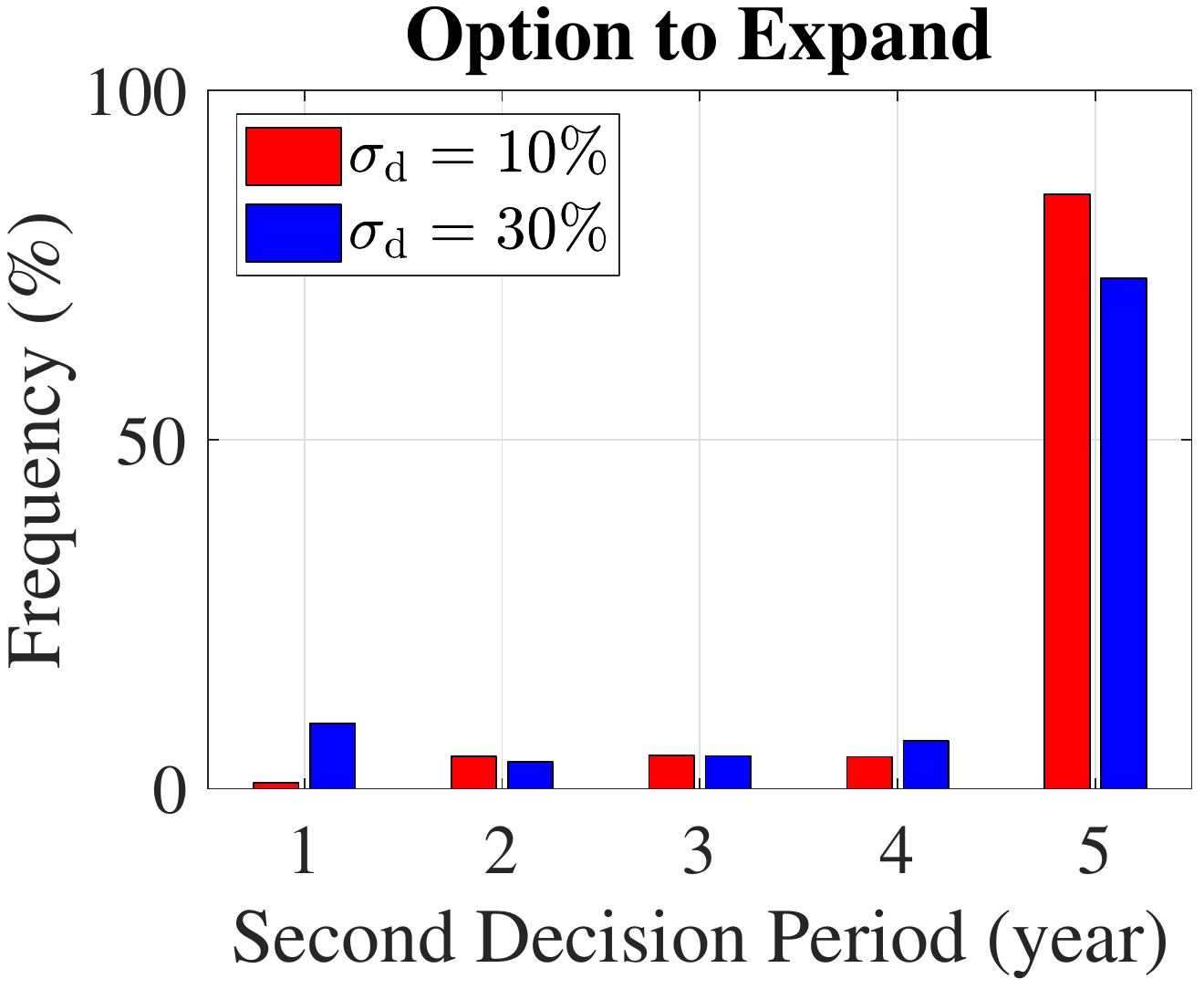}%
    \qquad
    \includegraphics[width=4cm,keepaspectratio]{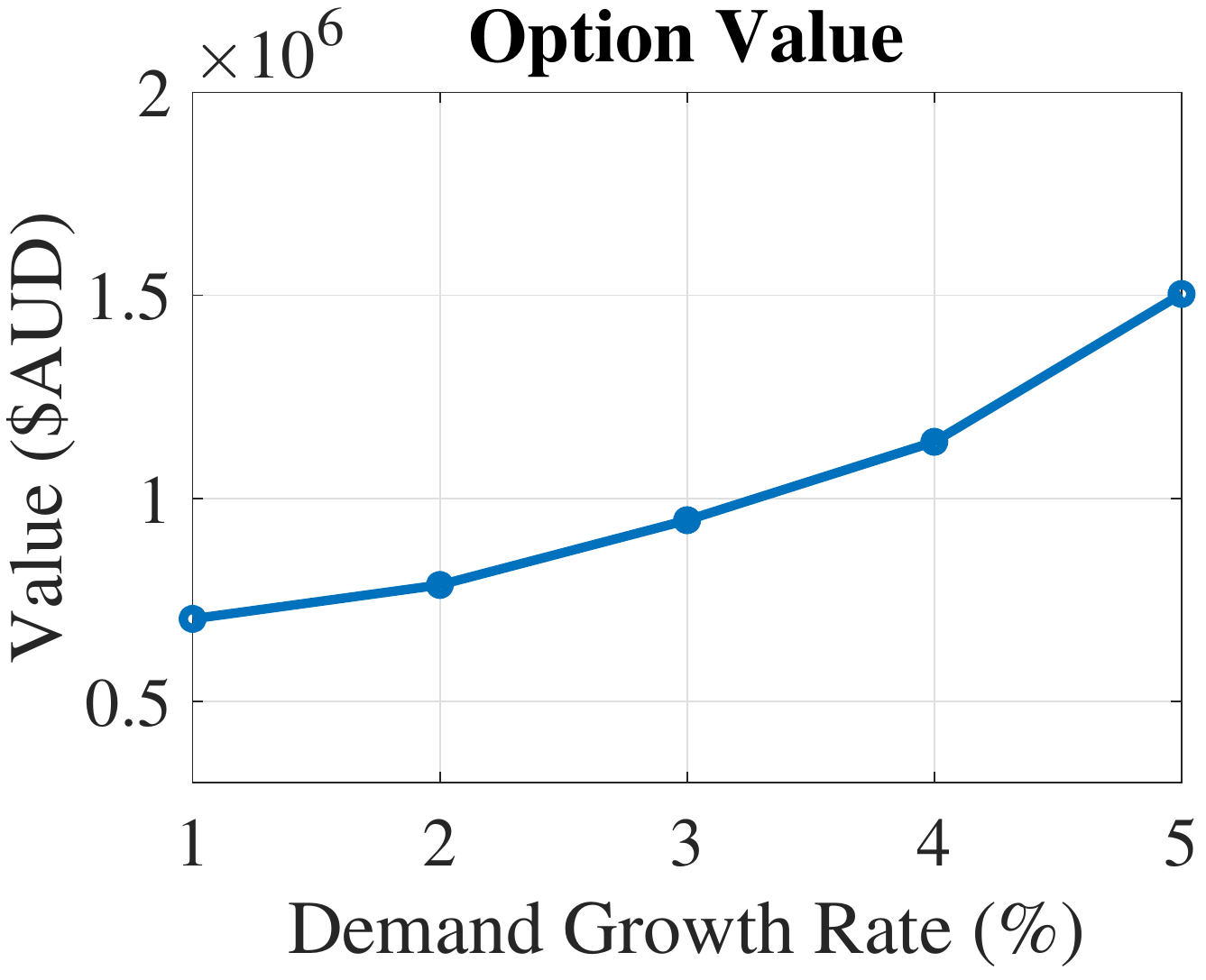}%
    \qquad
    \includegraphics[width=4cm,keepaspectratio]{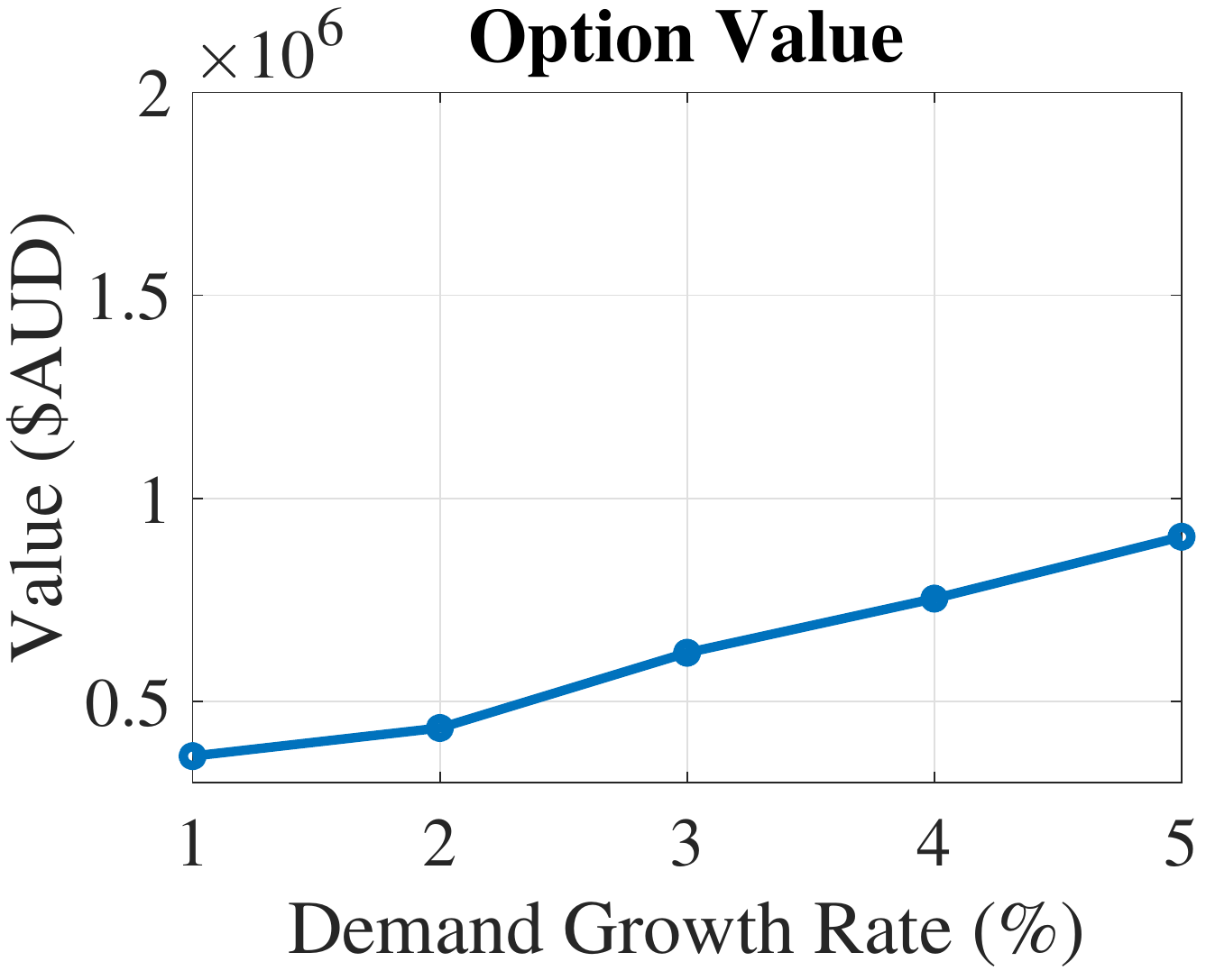}%
    \qquad
    \includegraphics[width=4cm,keepaspectratio]{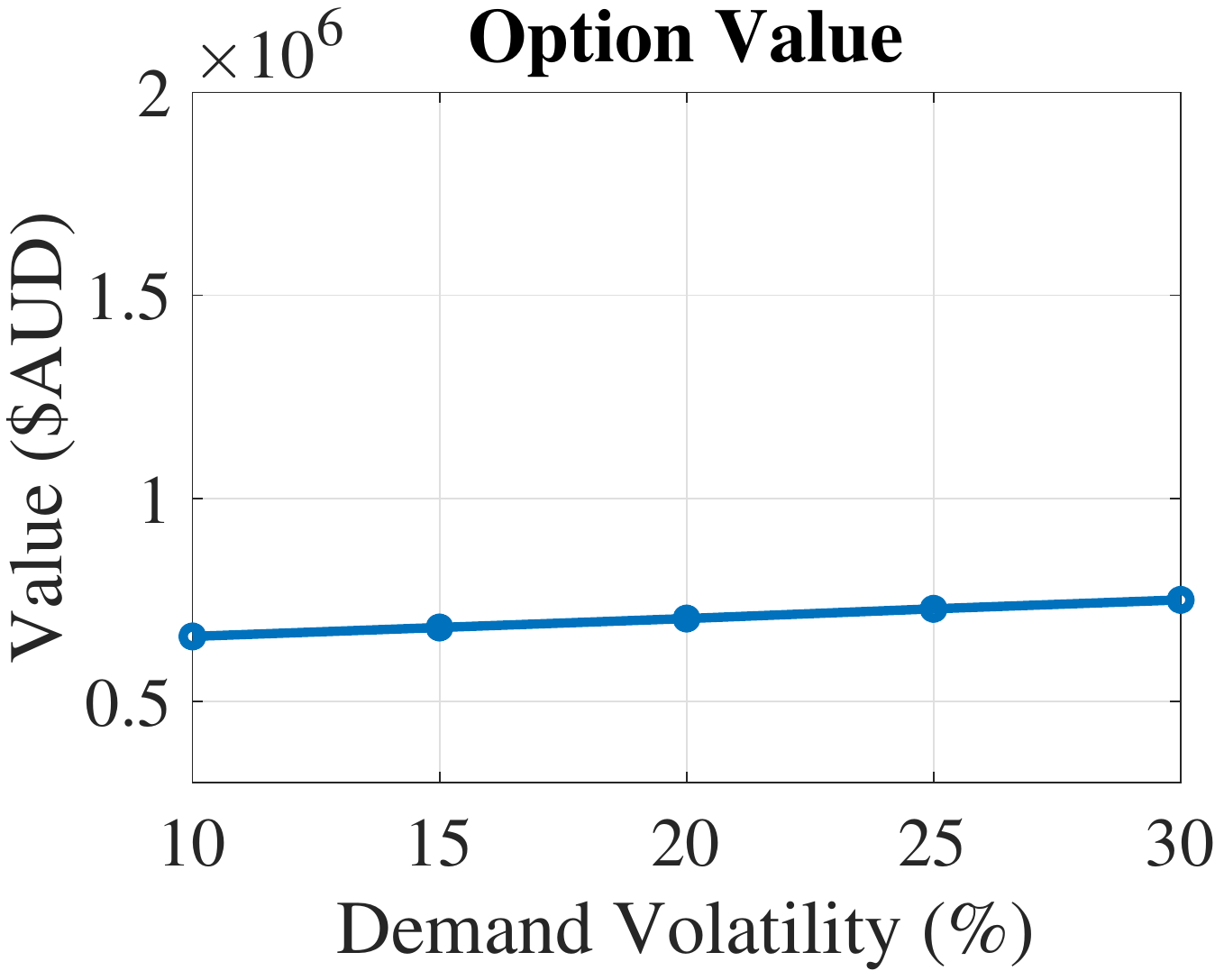}%
    \qquad
    \includegraphics[width=4cm,keepaspectratio]{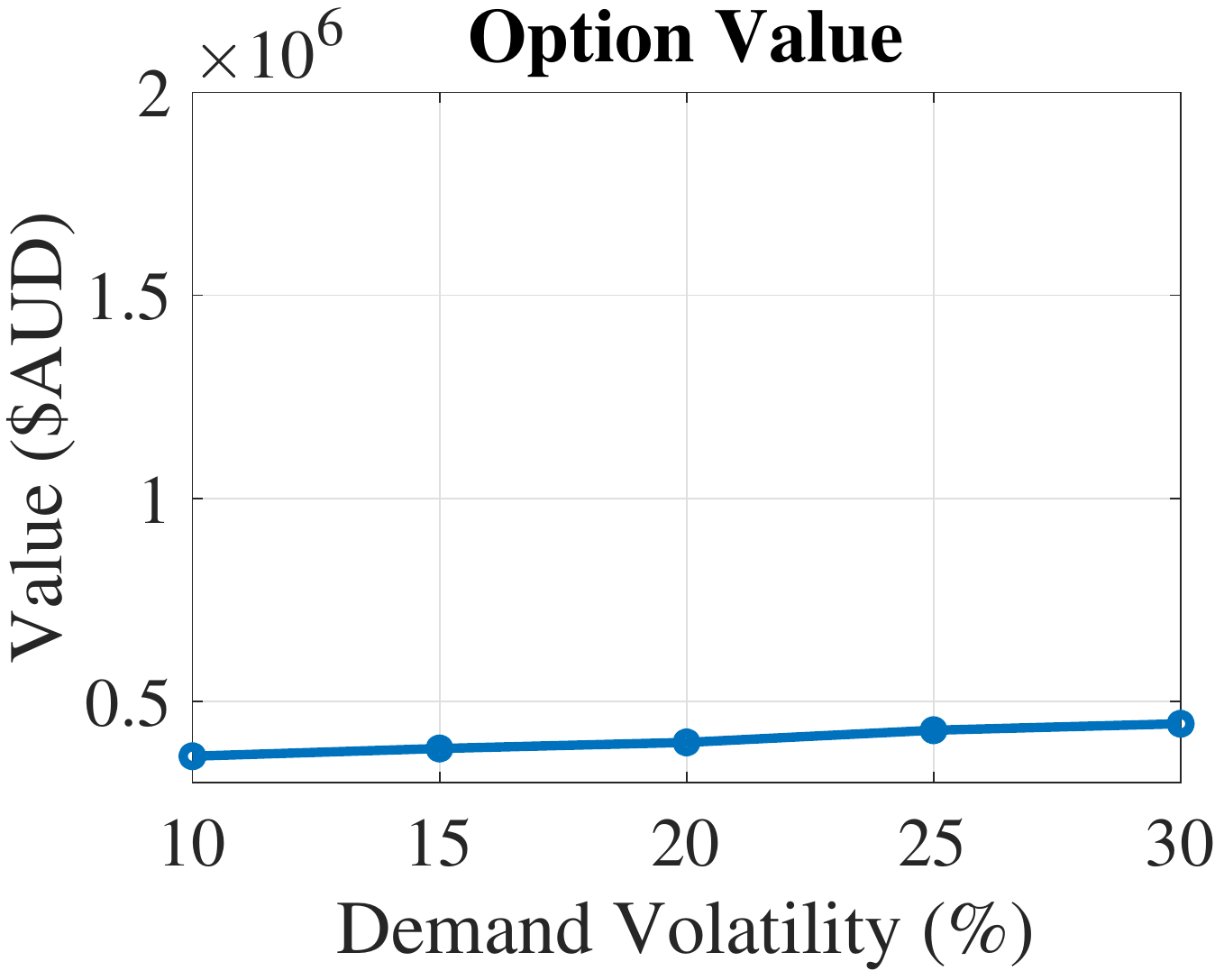}%
	\caption{Impact on optimal investment timing with respect to changing drift parameter and volatility of the state variables.}
    \label{fig5}%
\end{figure*}

\subsection{Real Options Valuation}
The opportunity values provided by future uncertainties are calculated using the proposed ROV framework. Specifically, the ROV suggests to delay the investment and wait for the market conditions to turn favourable if the deferral option value is positive and greater than the payoff calculated via the NPV method, and the optimal investment timing is when the payoff exceeds this option value. The flexible investment value is equal to the sum of the option value and the payoff calculated by the NPV analysis.

The LSMC approach for calculating the value of the compound options and the corresponding optimal investment strategies are detailed in Section \ref{section2}: (i) use the payoff (${\Pi_{h+1,t,\omega}}$) of expanding P2 to calculate the continuation value, ${\Phi_{h+1,t,\omega}}$, via least square regression, (ii) compare the payoff with continuation value for each year during the decision period using \eqref{ROVeq6}, the optimal timing to execute the investment on the $\omega^{\mathrm{th}}$ path is when the payoff is greater than the continuation value for the first time, (iii) calculate the investment value considering managerial flexibility using \eqref{ROVeq7}, and (iv) incorporate the value of the option to expand, $F_{h+1}$, in the LSMC method to determine the deferral option value, $F_{h}$, and the optimal investment strategy. The options are abandoned in the case of a negative payoff to reduce the computation time. 

The average payoff of the investment from the standard NPV analysis is {--\$}{240}{k}, as observed in Fig.~\ref{fig3}, Investment. The value of the compound options calculated using \eqref{ROVeq7} is {\$}{660}{k}. Therefore, the flexible investment value in this case equals {\$}{420}{k} ({\$}{660}{k} + (-- {\$}{240}{k})), as indicated in Table.~\ref{T2}, Benchmark. In this scenario, the drift parameter and volatility of the state variables are given in Table.~\ref{T1}. Based on the standard NPV approach, the investment is abandoned immediately because the payoff is negative. However, ROV sees potential hidden within the state variables, and discovers a possible additional benefit of {\$}{660}{k} if the investment is postponed to a later year. To predict the optimal investment timing, the LSMC computes the optimal stopping timing (${\tau_{\omega}}$) for each MC path, from this we extract the frequency distribution of ${\tau_{\omega}}$, for the deferral option, as shown in Fig.~\ref{fig4}, in red. The optimal timing of the investment is indicated by the greatest frequency. Thus, P2 is delayed to Year 5 (90\%), despite the fact that the DNSP needs to pay for the additional grid supply that covers the future growth in peak demand, $c^{\mathrm{g}}_{t,\omega}$, incurred by postponing P2. 

On the other hand, the option to expand calculated using \eqref{ROVeq7} in the second decision period is worth {\$}{330}{k}, while the payoff is only {\$}{30}{k} in Year 6, as seen in Fig.~\ref{fig3}, Expansion. In this case, the investment value is already positive before considering the flexibility, thus, based on the NPV analysis, the investment is expanded immediately. However, the ROV sees the potential for the market conditions to become favourable in the future, and hence suggests to delay the expansion. Specifically, based on the frequency distribution extracted from LSMC, as shown in Fig.~\ref{fig4}, blue, the option to expand is to be executed in Year 5 in the second decision period, with a 81\% possibility.     

If the deferral option is considered independent of the option to expand, the deferral option value is reduced from {\$}{660}{k} to only {\$}{350}{k}, while the optimal investment timing is kept in Year 5 with a slight decrease in the frequency (84\%), compared to Fig.~\ref{fig4}. Eliminating the option to expand in this calculation means that there is less managerial flexibility considered, leading to a smaller option value. This result shows that properly establish the relation between interacting options leads to a greater option value, and hence significantly increases the investment value. 

To summarize, the ROV framework accounts for the opportunity values from executing options under future uncertainties, and properly considers the interaction between these options. By doing so, it reverses the decisions drawn from the traditional NPV analysis. Specifically, the fast-declining PV-battery cost is the main driver to the deferral of the investment. This case serves as the benchmark in our work. 

\subsection{Sensitivity Analysis} 
The optimal investment strategy can be affected by altering the state variables, including growing power demand, varying diesel fuel price and the declining costs of PV-battery systems. Thus, it is useful to evaluate the sensitivity of the option value and optimal investment timing subject to changing scenarios characterized by these variables. The ROV results are summarized in Table.~\ref{T2}.

Overall speaking, increasing the growth rate of future peak demand (S2) leads to a greater operational cost for the diesel generator, and thus a greater standard NPV if the PV-battery investment is executed in Year 1. Meanwhile, this also means a greater deferral option value, because the payoff increases further by delaying the investment. On the other hand, increasing the volatility of future peak demand shifts the optimal investment timing for some paths to early years, as seen in Table.~\ref{T2}, S3. Nevertheless, the option value rises due to the increasing potential for a greater payoff in the future, compared to the benchmark. Increasing the mean reversion speed $\beta_{\mathrm{f}}$ to 15\% means diesel fuel price will reach the reversion level faster. As a result, the market becomes more confident to defer to the investment as the probability to execute the investment in Year 5 nearly reaches 100\%. 

In more detail, we use the example of peak demand growth to demonstrate how changing drift parameter and volatility affect the optimal investment timing. Specifically, the optimal investment timing is kept in Year 5, while the frequency decreases from 97\% to just under 50\% when $\mu_{\mathrm{d}}$ rises from $1\%$ to $5\%$, as shown in Fig.~\ref{fig5}, column one, top. As the demand grows faster, greater payoffs are expected. Thus, there is a greater probability for the payoff to exceed the continuation value in early investment years, and we are less confident to execute the investment in Year 5. Further, we observe an increasing deferral option value as there is larger profit in the future with a greater demand growth rate (Fig.~\ref{fig5}, column one, bottom). Similarly, the value of the option to expand rises with the same increase in $\mu_{\mathrm{d}}$, while the investment frequency in Year 5 is decreased, as shown in Fig.~\ref{fig5}, column two. 
  
In contrast, as we increase the volatility of demand ($\sigma_{\mathrm{d}}$) from 10\% to 30\%, greater payoffs are expected, so are the amount of negative payoff cash flows (due to the increasing uncertainties). Given that all negative cash flows are discarded by the LSMC approach, overall we observe a subtle increasing trend for the option value (Fig.~\ref{fig5}, column three, bottom). However, even though both the NPV and option value are increasing, the probability to execute P2 has been reduced from 90\% to 77\% in Year 5 (Fig.~\ref{fig5}, column three, top) as there is a greater possibility for the payoffs to become negative. Due to the same reasons, similar changes are observed in the case of expanding the investment, as shown in Fig.~\ref{fig5}, column four. 

Therefore, the optimal investment timing and the option value react differently to different values of drift parameter and volatility of the state variables. Interestingly, a greater deferral option value does not necessarily mean that the investment should be further delayed. Our results show that the optimal investment timing depends on how payoff and continuation value interact with each other. Meanwhile, it is possible for the option value to decrease as we increase future uncertainties, because there will be a greater number of payoffs becoming negative as time moves forward. These payoffs are discarded when calculating the option value. The sensitivity analysis shows how ROV responds to uncertain events in electricity market, and hence increases the robustness of the decision making process. In general, it is up to the DNSP to check how ROV responds to changing state variables, re-evaluate the investment, and make contingent decisions as time moves forward.  

\section{Conclusions} \label{section5}
This work proposed a ROV framework that evaluates a distribution network investment that embeds compound options, and derives the optimal investment strategy under uncertain market environment. We demonstrated the characteristics of the framework by determining the optimal strategy for the investment in residential PV-battery systems, and calculating the values of the option to expand, and then the deferral option via the LSMC approach. Through the proposed ROV framework, we demonstrated that delaying the PV-battery investment/expansion to a later year when future uncertainties have turned favourable increases the investment value and mitigates the risk of financial losses. Meanwhile, the framework has shown that the investment value can be significantly increased when the interaction between the real options are properly considered. More importantly, the framework allows the managerial flexibility to optimally respond to uncertain events characterized by the underlying state variables. Thus, the proposed ROV framework can be employed in the future to value a distribution network investment with multiple interacting options, and derive the corresponding optimal investment strategy. In future research, we expect to include network uncertainties including the size and location of PV-battery systems, load behaviour in the ROV framework. Capturing value of the opportunities presented by these uncertainties requires solving the power flow problem within the MC analysis underpinned by the ROV.

\section*{References}

\bibliography{bibfile}

\end{document}